\documentclass[prb,aps,twocolumn,showpacs,amsmath,amssymb,floatfix,eqsecnum,superscriptaddress]{revtex4-1}

\usepackage{graphicx}
\usepackage{caption}
\usepackage{subcaption}
\usepackage{dcolumn}
\usepackage{bm}

\usepackage{braket}
\usepackage{mathrsfs}
\usepackage{verbatim}

\usepackage{wasysym}
\usepackage{booktabs}
\usepackage{slashed}

\usepackage{graphicx}
\usepackage{psfrag} 
\usepackage{colordvi} 
\usepackage{setspace}
\usepackage{caption}
\usepackage{subcaption}

\usepackage{braket}
\usepackage{wrapfig}

\newcommand\beq{\begin{equation} \begin{aligned}}
\newcommand\eeq{\end{aligned} \end{equation}}

\newcommand\bmb{\left( \begin{matrix}}
\newcommand\emb{\end{matrix} \right)}
\newcommand\gammamu{\gamma^{\mu}}
\newcommand\Dmu{D_{\mu}}
\newcommand\Amu{A_{\mu}}
\newcommand\gammaz{\gamma_0}
\newcommand\gammao{\gamma_1}
\newcommand\gammat{\gamma_2}

\newcommand\bE{{\bf E}}

\newcommand\Aft{A^{f2}}
\newcommand\Afth{A^{f3}}

\newcommand\Aoft{A_1^{f2}}
\newcommand\Aofth{A_1^{f3}}

\newcommand\Atft{A_2^{f2}}
\newcommand\Atfth{A_2^{f3}}

\newcommand\Athft{A_3^{f2}}
\newcommand\Athfth{A_3^{f3}}

\newcommand\Azfzo{A_0^{f01}}
\newcommand\Azfoz{A_0^{f10}}
\newcommand\Azfo{A_0^{f1}}
\newcommand\Azft{A_0^{f2}}
\newcommand\Azfth{A_0^{f3}}

\newcommand\Psib{\overline{\Psi}}
\newcommand\jb{\overline{j}}
\newcommand\Jb{\overline{J}}

\newcommand\bA{\bar{A}}

\newcommand\vx{\vec{x}}

\newcommand\Mb{\overline{M}}
\newcommand\Sb{\overline{S}}

\newcommand\spartial{\slashed{\partial}}
\newcommand\ps{\slashed{p}}

\newcommand\sA{\slashed{A}}

\newcommand\inv{^{-1}}
\newcommand\sxy{\sigma^s_{xy}}
\newcommand\emnl{\epsilon^{\mu\nu\lambda}}
\newcommand\lb{\left(}
\newcommand\rb{\right)}
\newcommand\lcb{\left\{}
\newcommand\rcb{\right\}}

\begin{document}

\title{Chiral spin liquids on the kagome Lattice}

\author{Krishna Kumar}
\affiliation{Department of Physics and Institute for Condensed Matter Theory, University of Illinois at Urbana-Champaign, 1110 West Green Street, Urbana, Illinois, 61801-3080, USA}

\author{Kai Sun}
\affiliation{Department of Physics, Randall Laboratory, University of Michigan, Ann-Arbor, Michigan 48109, USA}

\author{Eduardo Fradkin}
\affiliation{Department of Physics and Institute for Condensed Matter Theory, University of Illinois at Urbana-Champaign, 1110 West Green Street, Urbana, Illinois, 61801-3080, USA}
\affiliation{Kavli Institute for Theoretical Physics, University of California Santa Barbara, Santa Barbara, California 93106-4030, USA}
\affiliation{Departamento de F{\'\i}sica, Facultad de Ciencias Exactas y Naturales, Universidad de Buenos Aires, Ciudad Universitaria, Pabellon I, 1428 Buenos Aires, Argentina}

\date{\today}

\begin{abstract}
We study the nearest neighbor $XXZ$ Heisenberg quantum antiferromagnet  on the  kagome lattice. Here we consider the effects of several perturbations: a)  a chirality term, b) a Dzyaloshinski-Moriya term, and c) a ring-exchange type term on the bowties of the kagome lattice, and inquire if they can support
 chiral spin liquids as ground states. The method used to study these Hamiltonians is a flux attachment transformation that maps the spins on the lattice to fermions coupled to a Chern-Simons gauge field on the kagome lattice. This  transformation requires us to consistently define a Chern-Simons term on the kagome lattice. We find that the chirality term leads to a chiral spin liquid even in the absence of an uniform magnetic field, with an effective spin Hall conductance of $\sxy = \frac{1}{2}$ in the regime of $XY$ anisotropy. The Dzyaloshinski-Moriya term also leads a similar chiral spin liquid but only when this term is not too strong. An external magnetic field also has the possibility of giving rise to additional plateaus which also behave like chiral spin liquids in the $XY$ regime. Finally, we consider the effects of a ring-exchange term and find that, provided its coupling constant is large enough, it may trigger a phase transition into a chiral spin liquid by the spontaneous breaking of time-reversal invariance.
\end{abstract}

\maketitle

\section{Introduction}

Quantum Heisenberg antiferromagnets on two-dimensional  kagome lattices are strongly frustrated quantum spin systems making them an ideal candidate to look for exotic spin liquid type states. 
This model has been the focus of many theoretical and numerical efforts for quite some time and,  in spite of these efforts,  many of its main properties remain only poorly understood.  On a parallel track, there has also been  significant progress on the experimental side with the discovery of materials such as Volborthite and Herbertsmithite whose structures are  closely represented by the Heisenberg antiferromagnet on the kagome lattice. An extensive analysis of the present experimental and theoretical status of this problem can be found in a recent review by Balents.\cite{Balents2010}

It is well known that frustrated antiferromagnets can give rise to magnetization plateaus in the presence of an external magnetic field. Although the nature of these plateaus depend on the type of model as well as the regime under study, in two dimensions these states are  expected to be topological phases of the chiral spin liquid type in the $XY$ regime.\cite{Misguich2001} Previous works have also shown that in the Ising regime these magnetization plateaus behave like valence bond crystals (VBC).\cite{Cabra2004, Nishimoto2013} In a recent work, the same authors studied the nearest neighbor 
XXZ Heisenberg model on the kagome lattice using a newly developed flux attachment transformation that maps the spins which are hard-core bosons to fermions coupled to Chern-Simons gauge field.\cite{Kumar2014} Here we showed that in the presence of an external magnetic field, this model gives rise to magnetization plateaus at magnetizations $m = \frac{1}{3}, \frac{2}{3}$ and $\frac{5}{9}$ in the $XY$ regime. The plateaus at $m=\frac{1}{3}$ and $m=\frac{2}{3}$ behave like a Laughlin fractional quantum Hall state of bosons with an effective spin Hall conductance of $\sxy = \frac{1}{2}$ whereas the plateau at $\frac{5}{9}$ was equivalent to a Jain state with $\sxy = \frac{2}{3}$. 

In spite of the considerable work done on this problem since the early 1990s, the kagome  Heisenberg antiferromagnet without an external magnetic field is less well understood. Density matrix renormalization group (DMRG) studies (which crucially also included second and third neighbor neighbor antiferromangnetic exchange interactions) have found a topological phase in the universality class of the $\mathbb{Z}_2$ spin liquid.\cite{Yan2011,Jiang2012,Depenbrock2012} A $\mathbb{Z}_2$ spin liquid has been proposed by Wang and Vishwanath\cite{Wang2006} (using slave boson methods), and  by Fisher, Balents and Girvin\cite{Balents2002,Isakov2011}, in generalized  ferromagnetic $XY$ model with ring-exchange interactions. Similarly, a $\mathbb{Z}_2$ spin liquid was shown to be the ground state for the kagome antiferromagnet in the quantum dimer approximation.\cite{Misguich-2002} We should note, however, that an entanglement renormalization group calculation\cite{Evenbly2010} appears to favor a complex 36 site VBC as the ground state of the isotropic antiferromagnet on the kagome lattice.

A more complex phase diagram (which also includes a chiral spin liquid phase as well as VBC, as well as  N{\'e}el and non-colinear antiferromagnetic phases) was subsequently found in the same extension of the kagome antiferromagnet by Gong and coworkers.\cite{Gong2015} It has also been suggested\cite{He2014} that a model of the kagome antiferromagnet with second and third neighbor {\em Ising} interactions may harbor a chiral spin liquid phase. Still,  variational wave function studies have also suggested a possible $U(1)$ Dirac spin liquid state,\cite{Ran2007, Iqbal2011} but this is not consistent with the DMRG results. Bieri and coworkers\cite{Bieri2015} used variational wave function methods to study a  kagome system with ferro and antiferromagnetic interactions, and find evidence for a  gapless chiral spin liquid state. Other studies that used non-linear spin-wave theory\cite{Chernyshev-2014,Gotze-2015} had predicted a quantum disordered phase for spin-1/2 kagome antiferromagnets with $XY$ anisotropy.

The most recent DMRG studies on the plain (without further neighbors)  kagome Heisenberg antiferromagnet give a strong indication of a gapped time-reversal invariant ground state in the $\mathbb{Z}_2$ topological class.\cite{Depenbrock2012} As it turns out there are two possible candidates $\mathbb{Z}_2$ spin liquids: the double Chern-Simons theory\cite{Freedman-2004} with a diagonal $K$-matrix, $K=\textrm{diag}(2,-2)$, and the Kitaev toric code\cite{Kitaev-2003} (which is equivalent to the deconfined phase of the Ising gauge theory\cite{Fradkin-1979}). An important recent result by Zaletel and Vishwanath\cite{Zaletel-2015} proves that the double Chern-Simons gauge theory is not allowed for a system with translation and time-reversal invariance such as the kagome antiferromagnet, which leaves the $\mathbb{Z}_2$ gauge theory as the only viable candidate.

In a recent recent and insightful work, Bauer and coworkers\cite{Bauer-2014} considered a kagome antiferromagnet with a term proportional to the chiral operator on each of the triangles of the kagome lattice. In this model time-reversal invariance is broken explicitly. These authors used a combination of DMRG numerical methods and analytic arguments to show that, at least if the time-reversal symmetry breaking term is strong enough, the ground state is a chiral spin liquid state in the same topological class as the Laughlin state for bosons at filling fraction $1/2$. This state was also found by us\cite{Kumar2014} in the $1/3$ and $2/3$ magnetization plateaus. A more complex phase diagram was recently found in a model that also included second and third neighbor exchange interactions.\cite{Wietek2015} On a separate track, Nielsen, Cirac and Sierra\cite{Nielsen2012} constructed lattice models with long range interactions with ground states closely related to the $\nu=1/2$ Laughlin wave function for bosons and, subsequently, deformed these models to systems with short range interactions on the square lattice that, in addition to first and second neighbor exchange interactions, have chiral three-spin interactions on the elementary plaquettes, and showed (using finite-size diagonalizations on small systems) that the ground state is indeed a chiral spin liquid.\cite{Nielsen2013} 

In this paper we investigate the occurrence of chiral spin liquid states in three extensions of the quantum Heisenberg antiferromagnet on the kagome lattice: a) by adding the chiral operator acting on the triangles of this lattice, b) by considering the effects of a Dzyaloshinski-Moriya interaction (both of which break time reversal invariance explicitly), and c) by adding a ring-exchange term on the bowties of the kagome lattice. This  term, equivalent to the product of two chiral operators of the two triangles of the bowtie, does not break time-reversal invariance explicitly, and  allows us to investigate the possible spontaneous breaking of time-reversal. Throughout we will use a flux-attachment procedure suited for the kagome lattice that we developed recently.\cite{Kumar2014,Sun2014} This method is well suited to investigate  chiral spin liquid phases but  not as suitable for  $\mathbb{Z}_2$ phases (at least not  straightforwardly). 

Throughout this paper we use a flux attachment transformation that maps the spins on the kagome lattice to fermions coupled to a Chern-Simons gauge field.\cite{Fradkin1989} The flux attachment transformation requires us to rigorously define a Chern-Simons term on the lattice. Previously, such a lattice Chern-Simons term had only been written down for the case of the square lattice.\cite{Eliezer1992a, Eliezer1992}  More recently, we have shown how to write down a lattice version of the Chern-Simons term for a large class of planar lattices.\cite{Sun2014} Equipped with these new tools, we can now study the nearest neighbor Heisenberg Hamiltonian, chirality terms and the Dzyaloshinski-Moriya terms on the kagome lattice, as well as ring exchange terms. Furthermore, this procedure will allow us to go beyond the mean-field level and consider the effects of fluctuations which are generally strong in frustrated systems. 

In spite of the successes of these methods, we should point out some of its present limitations. In a separate publication,\cite{Sun2014} we showed that the discretized construction of the Chern-Simons gauge theory (on which we rely heavily) can only be done consistently on a class of planar lattices for which there is a one-to-one correspondence between the sites (vertices) of the lattice and its plaquettes. At present time this restriction does not allow us to use these methods to more general lattices of interest (e.g. triangular). 

We should also
stress that in this approach there is no small parameter to control the accuracy of the approximations that are made. As it is well known from the history of the application of similar methods, e.g. to the fractional quantum Hall effects,\cite{Kalmeyer1987,Zhang-1989,Jain1989,Lopez1991,Yang1993}  that they can successfully predict the existence stable of chiral phases provided the resulting state has a gap already at the mean field level. These methods  predict correctly the universal topological properties of these topological phases,  in the form of effective low-energy actions, which encode  the correct form of the topologically-protected responses as well as the large-scale entanglement properties of the topological phases.\cite{Wen-1995,Kitaev-2006a,Levin-2006,Dong-2008,Zhang2012,Grover2013} However, these methods cannot  predict with significant accuracy the magnitude of dimensionful quantities such as the size of gaps since there is no small parameter to control the expansion away from the mean field theory. Thus, to prove that the predicted phases do exist for a specific model  is a motivation for further (numerical) studies. In particular, the recent numerical results of Bauer and coworkers\cite{Bauer-2014} are consistent with the results that we present in this paper.

Using these methods, we find that chiral spin liquid phases in the same topological class as the Laughlin state for bosons at filling fraction $1/2$ occur for both the model with the chiral operator on the triangles  and for the Dzyaloshinski-Moriya interaction. We find that the chirality term opens up a gap in the spectrum and leads to a chiral spin liquid state with an effective spin Hall conductivity of $\sigma_{xy}^s = \frac{1}{2}$ in the $XY$ regime. This is equivalent to a Laughlin fractional quantum Hall state for bosons similar to the $m=\frac{1}{3}$ magnetization plateau found in our earlier work.\cite{Kumar2014} We also show that this chiral state survives for small values of the Dzyaloshinski-Moriya term. The main motivation for this study was motivated by a recent numerical work\cite{Bauer-2014} where the authors studied the nearest neighbor Heisenberg model in the presence of the chirality and Dzyaloshinski-Moriya terms and found evidence of a similar fractional quantum Hall state with filling fraction $\frac{1}{2}$. The results we obtain agree qualitatively with the results obtained in the numerical work.
Since this is the same state that we obtained at the $1/3$ and $2/3$ magnetization plateaus, we searched for more complex topological phases by also adding a magnetic field. 
We   also consider the effects of adding the chirality term and Dzyaloshinski-Moriya terms in the presence of an external magnetic field at the mean-field level. (For numerical studies of magnetization plateaus in the kagome antiferromagnets see the recent work by Capponi {\it et al.}\cite{Capponi2013}). Once again, this is expected to give rise to magnetization plateaus. In the $XY$ regime, we again find some of the same plateaus that were already obtained for just the case of the Heisenberg model.\cite{Kumar2014} In addition, we also find an additional plateau at $m= \frac{1}{9}$ which has an effective spin Hall conductivity of $\sigma^s_{xy} = \frac{2}{3}$.
Finally, in order to address the possible spontaneous breaking of time-reversal, we  consider the effects of  a ring-exchange term on the bowties of the kagome lattice. Provided that the coupling constant of this ring-exchange term is large enough, this term triggers the spontaneous breaking of  time-reversal symmetry and leads to a similar chiral spin liquid state with $\sigma^s_{xy} = \frac{1}{2}$.

The paper is organized as follows. In Sec. \ref{sec:flux-attachment-kagome}, we will explicitly write down the $XXZ$ Heisenberg Hamiltonian and the chirality terms that we will consider. We will then briefly review the flux attachment transformation that maps the hard-core bosons (spins) to fermions and write down the resultant Chern-Simons term on the kagome lattice. A more detailed and formal discussion on the flux attachment transformation and the issues related to defining a Chern-Simons term on the lattice is presented elsewhere.\cite{Sun2014} An explicit derivation of the Chern-Simons term for the case of the kagome lattice was also presented in our earlier work.\cite{Kumar2014} In Sec. \ref{sec:MFT}, we setup the mean-field expressions for the nearest neighbor Heisenberg model in the presence of a chirality term. We will then begin by analyzing the mean-field state in the absence of the chirality terms and analyze the states obtained in the $XY$ and Ising regimes in Sec. \ref{sec:MF_XXZ}. Then we will consider the effects of adding the chirality term in Sec. \ref{sec:MF_chiral}. At the mean-field level we will also study the effects of adding an external magnetic field to such a system by analyzing the Hofstadter spectrum in Sec. \ref{sec:MF_chiral_hext}. Next, we will repeat the same analysis but with the Dzyaloshinski-Moriya term added to the $XXZ$ Heisenberg model in Sec. \ref{sec:MF_DM}. In Sec. \ref{sec:Continuum}, we will return to the state discussed in Sec. \ref{sec:MF_chiral} and expand the mean-field state around the Dirac points and write down a continuum version of the action. From here we will systematically consider the effects of fluctuations and derive an effective continuum action. Finally, we will consider a model for spontaneous time-reversal symmetry breaking by adding, to the $XXZ$ Heisenberg antiferromagnetic Hamiltonian,  a ring exchange term on the bowties of the kagome lattice  in Sec \ref{sec:SSB}. In  Section \ref{sec:Conclusions} we summarize the key  results obtained in the paper, and discusses several open questions.

\section{Flux attachment on the kagome lattice}
\label{sec:flux-attachment-kagome}

In this section, we will briefly review the theory of flux attachment on the kagome lattice in the context of the $XXZ$ Heisenberg Hamiltonian \cite{Kumar2014}. This will set the stage for the use of the flux attachment transformation that we will use to study these models.


We will begin with the nearest neighbor $XXZ$ Heisenberg Hamiltonian in the presence of an external magnetic field

\beq
    H = & J\sum_{\langle i,j \rangle} \left\{ S^x_i S^x_j + S_i^yS_j^y + \lambda S_i^z S_j^z \right\} - h_{ext} \sum_i S_i^z
    \label{eq:Heisenberg_Hamiltonian}
\eeq
where $\langle i,j \rangle$ refer to the nearest neighbor sites on the kagome lattice and $\lambda$ is the anisotropy parameter along the $z$ direction. $h_{ext}$ refers to the external magnetic field.
Using the flux attachment transformation, the spins (which are hard-core bosons) can be mapped to a problem of fermions coupled to a Chern-Simons gauge field. The resultant action takes the form
\beq
S = S_F(\psi,\psi^*, \Amu) + S_{int}(\Amu) + \theta S_{CS}(\Amu)
\label{eq:Action_JW}
\eeq

\begin{figure}[h]
 \includegraphics[width=0.5\textwidth]{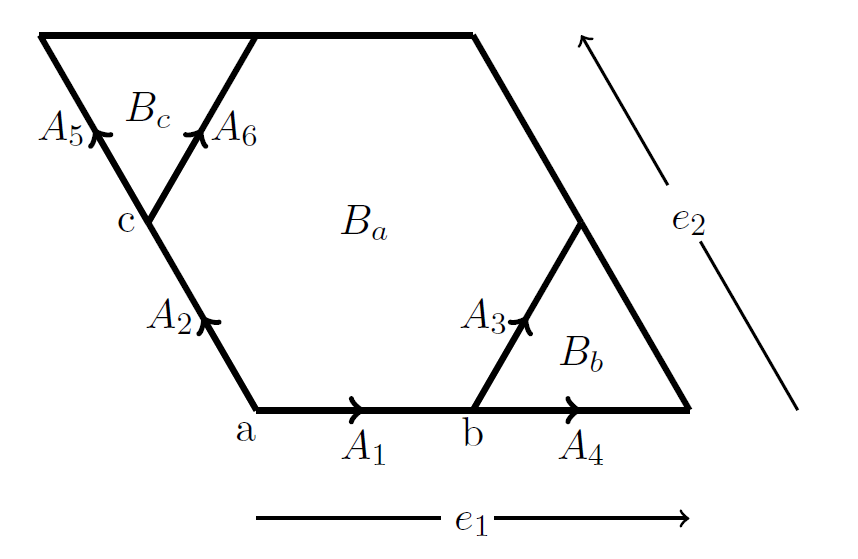}
\caption{Unit cell of the kagome lattice. The unit cell has three sub-lattice sites (labelled a, b and c) and three plaquettes (two triangles and one hexagon). The flux attachment transformation proceeds by attaching the fluxes in each of the plaquettes to its corresponding sites. }
\label{fig:kagome_unit_cell}
\end{figure}

The $S^xS^x$ and $S^yS^y$ terms map to the fermionic hopping part in the presence of the Chern-Simons gauge field, $A_j(x)$, and $S^z S^z$ terms map to fermionic interaction term as shown in the below equations
\beq
S_F(\psi,\psi^*,A_{\mu})  = & \int_t \sum_{{\bm x}} \bigg[ 					\psi^*(x)\left(iD_{0}+\mu\right)\psi(x)  \\
& -\frac{J}{2} \sum_{\langle \vx, \vx' \rangle} \left( \psi^*(x)e^{iA_j(x)}\psi(x')+h.c \right) \bigg] \\
S_{int}(\psi, \psi^*) = & \lambda J \int_t \sum_{\langle \vx,\vx' \rangle}\left(\frac{1}{2} - n(x) \right) \left( \frac{1}{2} - n(x') \right)
\label{eq:Action_fermions}
\eeq
where $D_0 = \partial_0 + i A_0$ is the covariant time derivative, $\langle \vx, \vx' \rangle$ stands for nearest neighbor sites $\vx$ and $\vx'$ on the kagome lattice and the space-time coordinate $x \equiv (\vx,t)$. The temporal gauge fields $A_0$ live on the sites of the kagome lattices and the spatial gauge fields $A_j(x)$ live on the links of the lattice as can be seen in the unit cell of the kagome lattice in Fig \ref{fig:kagome_unit_cell}.

The density operator $n(x) = \psi^*(x)\psi(x)$   is related to the $S^z$ spin component as follows
\beq
S^z(x) = \frac{1}{2} - n(x)
\label{eq:Sz_map}
\eeq
The above expression also allows us to absorb the external magnetic field term ($h_{ext}$) in to the definition of the chemical potential $\mu$, i.e. in the fermionic language the effect of the external magnetic field can be mimicked by changing the fermion density on the lattice. For a majority of this paper, we will focus on the case where $h_{ext} = 0$. This would correspond to the case of half-filling in the fermionic theory after the flux attachment transformation.

Now all that remains is the Chern-Simons term on the kagome lattice. An explicit derivation of this term for the case of the kagome lattice was already presented in an earlier paper.\citep{Kumar2014} A more detailed and rigorous representation of a Chern-Simons term on generic planar lattices is also presented elsewhere.\citep{Sun2014} Here, we will simply reproduce some of the relevant results required for our analysis.

The $\theta$ parameter in front of the the Chern-Simons term in Eq.\eqref{eq:Action_JW} is taken to be $\theta = \frac{1}{2\pi}$ to ensure that the statistics of the spins (hard-core bosons in Eq. \eqref{eq:Heisenberg_Hamiltonian}) are correctly transmuted to those of the fermions in Eq. \eqref{eq:Action_JW}. The Chern-Simons term on the kagome lattice can be written as

\beq 
	S_{CS} = & S_{CS}^{(1)} + S_{CS}^{(2)} \\
	S_{CS}^{(1)} & =  \int dt \sum_{x,y}  A_{0}(x,t)J_{i}(x-y)A_{i}(y,t)  \\	
	S_{CS}^{(2)} &=  - \frac{1}{2}\int dt \sum_{x,y}A_{i}(x,t)K_{ij}(x-y)\dot{A_{j}}(y,t)
	\label{eq:CS_action}
\eeq
The first term $S_{CS}^{(1)}$ in Eq.\eqref{eq:CS_action} is the flux attachment term that relates the density at a site on the lattice to the flux in its corresponding plaquette. For the case shown in Fig \ref{fig:kagome_unit_cell}, the explicit expression for this term is given as
\beq
	{\bm J}^a(k) & = (1, -1, 1, -e^{-i k_2}, e^{-i k_1}, 1) \\
	{\bm J}^b(k) & = (0, e^{-i k_1}, -1, 1, 0, 0) \\
	{\bm J}^c(k) & = (-e^{-i k_2}, 0, 0, 0, -1, 1) 
\eeq
where $k_1$ and $k_2$ are the Fourier components along the $e_1$ and $e_2$ directions of the unit cell shown in Fig \ref{fig:kagome_unit_cell}. These choices ensure that the fermion density $n(x)$ (at a site $x$ of the kagome lattice) is related to the gauge flux $B(x)$ on the adjoining plaquette by the constraint equation
$n(x)=\theta B(x)$
as an operator identity on the Hilbert space.

The second term in Eq.\eqref{eq:CS_action} establishes the commutation relations between the different gauge fields on the lattice and it is the structure of the $K_{ij}$ matrix that ensures that the fluxes commute on neighboring sites. This condition is crucial to being able to enforce the flux attachment constraint consistently on each and every site of the lattice. The explicit expression for the $K_{ij}$ matrix is given as
\begin{widetext}
\begin{equation}
K_{ij} = 
\frac{1}{2} \left(
\begin{array}{cccccc}
 0 & -1 & 1 & -s_2 & s_1+s_2^{-1} & -1+s_2^{-1} \\
 1 & 0 & 1-s_1^{-1} & -s_2-s_1^{-1} & s_1 & -1 \\
 -1 & s_1-1 & 0 & 1-s_2 & s_1 & -1 \\
 s_2^{-1} & s_1+s_2^{-1} & s_2^{-1}-1 & 0 & s_1 s_2^{-1} & s_2^{-1} \\
 -s_2-s_1^{-1} & -s_1^{-1} & -s_1^{-1} & -s_2s_1^{-1} & 0 & 1-s_1^{-1} \\
 1-s_2 & 1 & 1 & -s_2 & s_1-1 & 0 \\
\end{array}
\right)
	\label{eq:K_matrix_kagome}
\end{equation}
\end{widetext}
where $s_j$ are shift operators along the two different directions ($e_1$ and $e_2$) on the lattice i.e. $s_j f(x) = f(x+e_j)$ as shown in Fig \ref{fig:kagome_unit_cell}.

\section{The $XXZ$ model with a chirality breaking field}
\label{sec:chirality-breaking}

Next, we will consider the effects of adding a chirality breaking term to the Heisenberg Hamiltonian in Eq.\eqref{eq:Heisenberg_Hamiltonian}. A system of spin-1/2 degrees of freedom on the kagome lattice with a chirality breaking term as its Hamiltonian was considered recently by Bauer and coworkers.\cite{Bauer-2014} Using finite-size diagonalizations and DMRG calculations, combined with analytic arguments, these authors showed that the ground state of this system with an explicitly broken time-reversal invariance is a topological fluid in the universality class of the Laughlin state for bosons  at level 2 (or, equivalently, filling fraction 1/2). Here we will examine this problem (including the $XXZ$ Hamiltonian) and find that the ground state has indeed the same universal features found by Bauer and coworkers, and by us in the 1/3 plateau.\cite{Kumar2014}

The resultant Hamiltonian is given as
\beq
  H_{\rm tot} = H_{XXZ} + H_{\rm ch} - h_{\rm ext}\sum_{i}S_i^z
  \label{eq:Chiral_Hamiltonian}
\eeq
where $H_{XXZ}$ is the $XXZ$ Heisenberg Hamiltonian in \eqref{eq:Heisenberg_Hamiltonian}. The chirality breaking term is given by 
\beq
  H_{\rm ch} = h \sum_{\triangle} \chi_{ijk}(\triangle) = h \sum_{\triangle} {\bm S}_i\cdot \lb {\bm S}_j \times {\bm S}_k \rb
  \label{eq:Chiral_Term}
\eeq
where $\chi_{ijk}(\triangle)$ is the chirality of the three spins on each of the triangular plaquettes of the kagome lattice and the sum runs over all the triangles of the kagome lattice. Recall the important fact that each unit cell of the of the kagome lattice contains two triangles.

In order to use the flux attachment transformation, it is convenient to express the spin operators $S^x$ and $S^y$ in terms of the raising and lowering $S^+$ and $S^-$. As an example, one can re-write the chirality term on a triangular plaquette associated with site $b$ (shown in Fig \ref{fig:kagome_unit_cell}) as follows
\beq
  \chi_{b} =& {\bm S}_{a}  \cdot \left( {\bm S}_{c}  \times {\bm S}_{b} \right) \\
  =& \frac{i}{2} \bigg\{ -S_{a}^-S_{c}^+S_{b}^z+S_{a}^+S_{c}^-S_{b}^z \\
     	&+S_{a}^-S_{c}^zS_{b}^+-S_{a}^+S_{c}^zS_{b}^- \\
     	&-S_{a}^zS_{c}^-S_{b}^++S_{a}^z S_{c}^+S_{b}^- \bigg\}     
	\label{eq:chirality-expanded}
\eeq
where the subscripts  $a$, $b$ and $c$ label  the three corners of a triangular plaquette in Fig \ref{fig:kagome_unit_cell}.

As shown in Ref. [\onlinecite{Kumar2014}] (and summarized in Section \ref{sec:flux-attachment-kagome}),
the raising and lowering spin operators $S^\pm$ are interpreted as the creation and destruction operators for bosons with hard cores, and $S^z$ operators are simply related to the occupation number $n$ of the bosons by $S^z=\frac{1}{2}-n$. Under the flux attachment transformation, the hard core bosons are mapped onto a system of fermions coupled to Chern-Simons gauge fields (residing on the links of the kagome lattice). The boson  occupation number at a given site is mapped (as an operator identity) onto the gauge flux in the adjoining plaquette (in units of $2\pi$).

It is the straightforward to see that the chirality term gets mapped onto an additional hopping term on the links of the kagome lattice which carries a gauge as an extra phase factor on each link determined by the fermion density on the opposite site of the triangle. As a result, only the fermionic hopping part of the action in Eq.\eqref{eq:Action_fermions} gets modified, and the interaction part and the Chern-Simons part are unaffected. 

Putting things together we get an effective fermionic hopping part that has the form
\beq
S_F&(\psi,\psi^*,A_{\mu})  =  \int_t \sum_{{\bm x}} \bigg\{ 					\psi^*(x)\left(iD_{0}+\mu\right)\psi(x)  \\
-& \sum_{\langle \vx, \vx' \rangle}\Jb(x_{(a)}) \lb e^{-i \phi(x_{(a)})} \psi^*(x)e^{iA_j(x)}\psi(x')+h.c \rb \bigg\}
\label{eq:JW_chirality} 
\eeq
where once again $x$ and $x'$ are nearest neighbor sites and $x_{(a)}$ refers to third site on the triangle formed by sites $x$ and $x'$. The subscript $(a)$ refers to the sub-lattice label. The expressions of $\Jb$ and $\phi$ on each sub-lattice can be written as
\beq
\Jb_{(a)}(x) =& \frac{1}{2}\sqrt{ J^2 + h^2 \lb \frac{1}{2} - n_{(a)}(x) \rb^2 }	\\
\phi_{(a)}(x) =& \text{tan} \inv \left[ \frac{h}{J}\lb \frac{1}{2} - n_{(a)}(x) \rb \right]
\label{eq:Jb}
\eeq
Hence, we have expressed the effects of the chirality term in terms of a modified hopping strength $\Jb_{(a)}$ and an additional gauge field ($\phi_{(a)}(x)$) on each of the links of the lattice. In the limit that $h = 0$, we just have the original gauge fields and in the other limit with $J=0$ each link has an additional contribution of ($\phi_{(a)}(x) = \pm \frac{\pi}{2}$). 

\section{Mean-field theory}
\label{sec:MFT}

In this section, we will set up the mean-field expressions for the fermionic action in Eq.\eqref{eq:Action_fermions} and Eq.\eqref{eq:JW_chirality}. The basic setup here is very similar to the situation described in our earlier work,\cite{Kumar2014} but it has been modified to account for the addition of the chirality term in this paper.

Using the flux attachment constraint imposed by the Chern-Simons term ($n(x) = \theta B(x)$), the interaction term in Eq.\eqref{eq:Action_fermions} can now be re-written as follows 
\beq
	L_{\rm int}(A_{\mu}) = \lambda J \sum_{\langle \vx, \vx' \rangle}\left(\frac{1}{2} - \theta B(x) \right) \left( \frac{1}{2} - \theta B(x') \right)
	\label{eq:Interaction_A}
\eeq
The interaction term has been expressed purely in terms of gauge fields. Hence, the resultant action after the flux attachment transformation is quadratic in the fermionic fields. Integrating out the fermionic degrees of freedom gives rise to the below effective action just in terms of the gauge fields
\beq
S_{\rm eff}(\Amu)= -& i \textrm{tr}\ln[iD_{0}+\mu-H_{\rm hop}(A)] \\
	      & +S_{\rm int}(\Amu)+\theta S_{CS}(\Amu)
	\label{eq:Action_gauge_fields}
\eeq
where the hopping Hamiltonian $H_{\rm hop}(A)$ is (in matrix notation)
\beq
	H_{\rm hop}=\sum_{\langle \vx, \vx' \rangle}\left\{\Jb_{(a)} e^{i A_j(x)-i\phi_{(a)}} \ket{x} \bra{x'}+h.c\right\} 
	\label{eq:Hamiltonian_matrix}
\eeq
where the above sum runs over all nearest neighbors $\vx$ and $\vx'$. The gauge field $A_j(x)$ refers to the hopping term required to go from point $\vx$ to $\vx'$ on the lattice. The term $\Jb_{(a)}$ and $\phi_{(a)}$ are as defined in Eq.\eqref{eq:Jb} with $(a)$ once again referring to the sub-lattice index. In the above expression $(a)$ would correspond to the third site in the triangle formed by nearest-neighbor sites $\vx$ and $\vx'$.

Now the mean-field equations can be obtained by extremizing the action in Eq.\eqref{eq:Action_gauge_fields} w.r.t. the gauge fields 
\begin{equation}
\frac{\delta S_{\rm eff}(A)}{\delta A_{\mu}}\Bigg|_{A_{\mu}=\bar{A}_{\mu}}=0
\end{equation}
Differentiation with respect to the time components $A_0$ yields the usual equation relating the density to the flux,
\begin{equation}
	\langle n(x) \rangle=\frac{1}{2\pi}\langle B(x) \rangle
	\label{eq:MF_density}
\end{equation}
which implies that the flux attachment is now enforced at the mean-field level. The average density can be expressed in terms of  the mean-field propagator by
\begin{equation}
\langle n(x,t) \rangle=\left\langle -\frac{\delta S_{F}}{\delta A_{0}(x,t)}\right\rangle =-iS(x,t;x,t)
\end{equation}
where $S_F$ refers to just the fermionic part of the action (i.e. the hopping part) and $S(x,t;x',t')$ is the   fermion propagator in an average background field $\bar{A}_{\mu}(x,t)$. 

Differentiation with respect to the spatial $A_k$ components yields an expression for the local currents,
\begin{equation}
\left\langle j_{k}(x,t)\right\rangle =\theta\left\langle \frac{\delta S_{CS}}{\delta A_{k}(x,t)}\right\rangle +\left\langle \frac{\delta S_{\rm int}}{\delta A_{k}(x,t)}\right\rangle 
\label{eq:saddle_current}
\end{equation}
Here too, we can express the average current in terms of the fermionic action in the usual manner
\beq
\langle j_{k}(x,t) \rangle= &  \left\langle -\frac{\delta S_{F}}{\delta A_{k}(x,t)}\right\rangle 
\eeq

We will look for uniform and time-independent solutions af these equations. Under these conditions the mean-field equations for the currents, Eq.\eqref{eq:saddle_current}, becomes
\beq
\langle  j_{k}(x) \rangle =& 
 \theta \bar{d}^{k\alpha}\bar{A}_{0\alpha}(x) \\
	 &-2J\lambda\theta^{2} (-1)^{k} \left[ \bar{B}^{a} - f_{k} \bar{B}^{c} - (1-f_{k}) \bar{B}^{b} \right]
\label{eq:MF_current}
\eeq
with $f_k = 1$ when $k = 1, 5, 6$ and $f_k = 0$ when $k = 2, 3, 4$. In the above expression, we have also fixed the average fluxes on each sub-lattice (i.e. the fluxes on all sub-lattices of a particular type are the same),  $\alpha$ is the sub-lattice index, and
\beq
\bar{d}^{k\alpha} = 
\bmb
1 & 0 & -s_2^{-1} \\
-1 & s_1^{-1} & 0	\\
1 & -1 & 0 \\
-s_2^{-1} & 1 & 0 \\
s_1^{-1} & 0 & -1 \\
-1 & 0  & 1
\emb
\eeq
where $s_1$ and $s_2$ are the same shift operators discussed after Eq.\eqref{eq:K_matrix_kagome}.

\subsection{Mean-field ansatz for $XXZ$ model}
\label{sec:MF_XXZ}

\begin{figure}
  \begin{center}
    \includegraphics[width=0.48\textwidth]{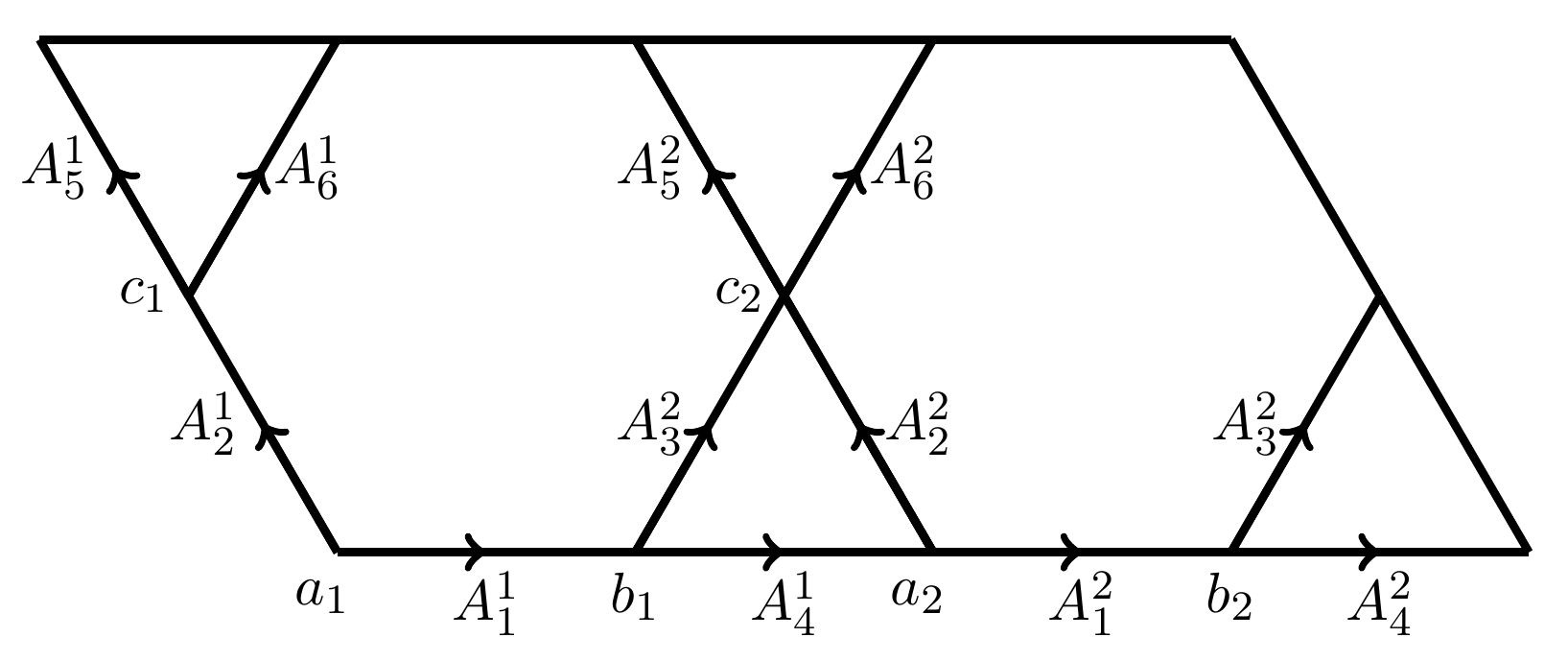}
  \end{center}
  \caption{Magnetic unit cell at half filling. a, b and c label the different sub-lattices in each of the unit cells. The gauge fields now have an additional label to indicate the unit-cell they belong to.}
  \label{fig:MF_half_filling}
\end{figure}

We will begin by studying the case with of the $XXZ$ Heisenberg model i.e. we set to zero both the chirality coupling $h$ and the external magnetic field, $h_{\rm ext}$. This translates to the case of half-filling in the fermionic language. At $\frac{1}{2}$ filling, the average density within each unit cell is given by
\beq
\frac{1}{3} \left(\langle n_a \rangle + \langle n_b \rangle + \langle n_c \rangle\right)= \frac{1}{2}
\eeq
 where $a$, $b$ and $c$ refer to the three sublattices. This gives an average flux of $\pi$ in each unit cell which implies that the magnetic unit cell consists of two unit cells as shown in Fig \ref{fig:MF_half_filling}.

In the absence of the chirality term we will primarily look for mean-field phases that are uniform and time-independent, and have zero currents, i.e. $\langle j_k(x,t) \rangle = 0$ in Eq.\eqref{eq:MF_density} and Eq.\eqref{eq:saddle_current}. The flux attachment condition can be imposed as follows on each of the sub-lattices
\beq
	\langle n_a(x) \rangle =& \theta \langle B_a(x) \rangle = \frac{1}{2} - \Delta_1 - \Delta_2 \\
	\langle n_b(x) \rangle =& \theta \langle B_b(x) \rangle = \frac{1}{2} + \Delta_1 \\
	\langle n_c(x) \rangle =& \theta \langle B_c(x) \rangle = \frac{1}{2} + \Delta_2 
  \label{eq:MF_flux_XXZ}
\eeq 
where $\Delta_1$ and $\Delta_2$ are two parameters that will be chosen to satisfy the mean-field self-consistency equations. 
The fluxes in Eq.\eqref{eq:MF_flux_XXZ} can be achieved by the below choice of gauge fields in Fig \ref{fig:MF_half_filling}
\beq
\setlength\extrarowheight{4pt}
\begin{matrix}
  A_1^1 = 0	& & \quad	A_1^2 = 0	\\
  A_2^1 = p_1	& & \quad	A_2^2 = p_1	\\
  A_3^1 = 0	& & \quad	A_3^2 = 0	\\
  A_4^1 = 0	& & \quad	A_4^2 = 0	\\
  A_5^1 = -p_2	& & \quad	A_5^2 = -p_2+3\pi	\\
  A_6^1 = 0	& & \quad	A_6^2 = 3\pi	
\end{matrix}
\eeq
where $p_1 = \pi + 2\pi \Delta_1$ and $p_2 = \pi + 2\pi \Delta_2$.

With these expressions for the densities, the mean-field equation (Eq.\eqref{eq:MF_current}) can be satisfied by the below choices for the temporal gauge fields
\beq
	A_{0,a} = 2\lambda (\Delta_1 + \Delta_2)	& & &
	A_{0,b} = -2\lambda \Delta_1	& 
	A_{0,c} = -2\lambda \Delta_2
	\label{eq:A0_MFT}
\eeq
Using this mean-field field setup, we find two regimes at the mean-field level. 

\subsubsection{XY regime}
\label{sec:XY_J}

\begin{figure}[h]
 \begin{center}
  \includegraphics[width=\linewidth]{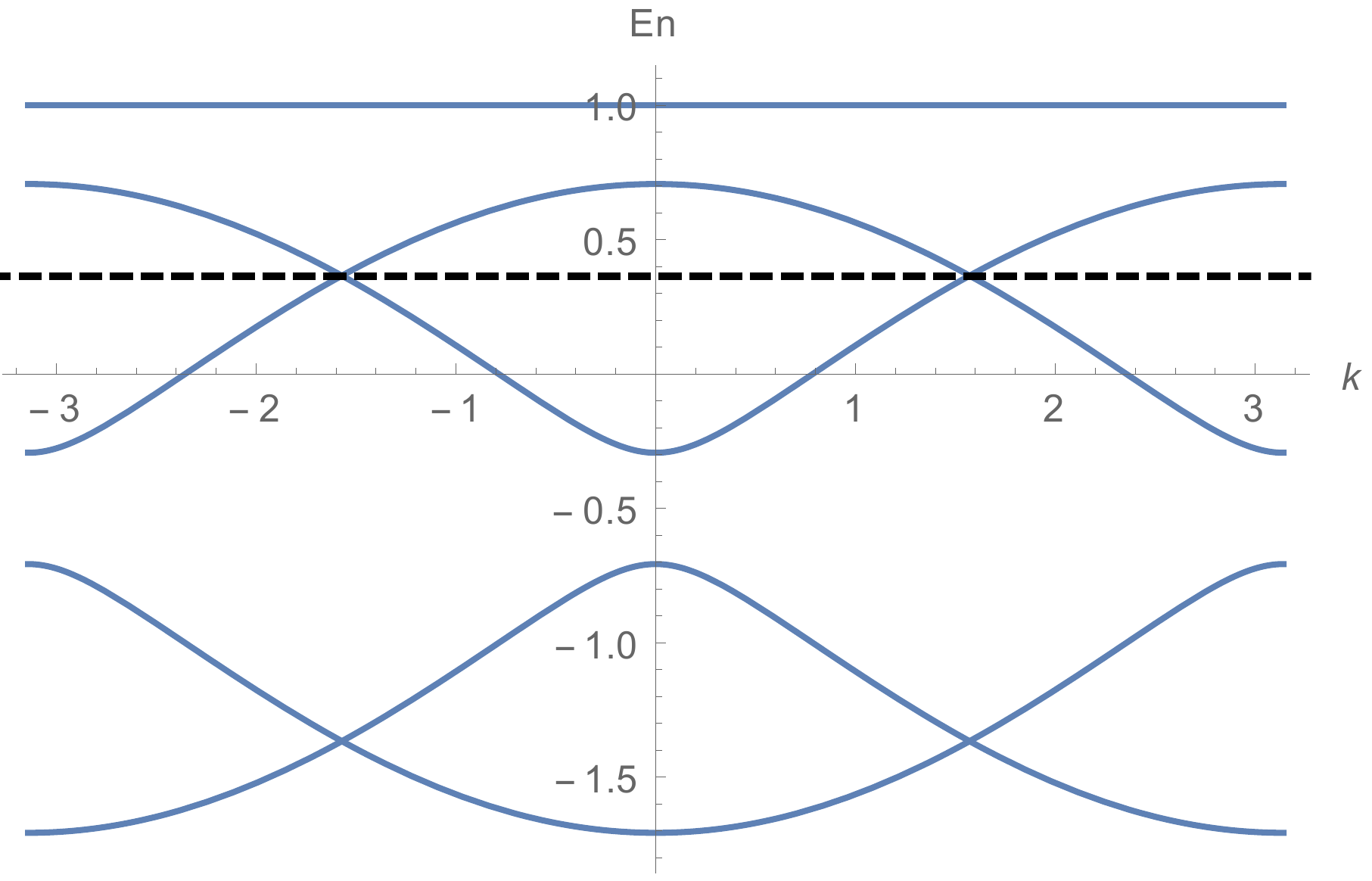}
  \end{center}
  \caption{(Color online) Mean-field spectrum in the $XY$ regime at half-filling, showing the two Dirac points. The dashed line indicates the Fermi level. The top band is doubly degenerate. These plots are made along the $k_y = -\frac{k_x}{\sqrt{3}}$ line in the Brillouin zone (along which the two Dirac points lie).}
  \label{fig:pi_flux_spectrum_Dirac}
\end{figure}

In the $XY$ regime, $\frac{\lambda}{J} \lesssim 1$, we find that $\Delta_1 = \Delta_2 = 0$ is the only solution that satisfies the self-consistency condition. This leads to a state with a flux of $\pi$ in each of the plaquettes. We will represent this as the $(\langle B_a \rangle, \langle B_b \rangle, \langle B_c \rangle)=(\pi, \pi,\pi)$ flux state. This state has a total of six bands, shown in Fig. \ref{fig:pi_flux_spectrum_Dirac} (the top two bands are double degenerate). At half-filling the bottom three bands are filled giving rise to two Dirac points in the spectrum, crossed by the dotted line in Fig.\ref{fig:pi_flux_spectrum_Dirac} which indicates the Fermi level. See Sec.\ref{sec:Continuum} for details.

At the mean-field level this spectrum is equivalent to the gapless $U(1)$ Dirac spin liquid state that has been discussed in previous works.\cite{Ran2007, Iqbal2011} We notice, however, that there are other works that  favor symmetry breaking states but with a doubled unit cell and a flux of $\pi$ in each of the plaquettes \cite{Clark2013}. The state we find could survive when fluctuations are considered giving rise to one of the above states. Alternatively, fluctuations could also open up a gap in the spectrum leading to an entirely different phase. In this paper, we will only analyze the gapless states at a mean-field level.

\subsubsection{Ising Regime}
\label{sec:Ising_J}

For $\frac{\lambda}{J} \gtrsim 1$, non-vanishing values of $\Delta_1$ and $\Delta_2$ are required to satisfy the mean-field consistency equations. The solution with the lowest  energy has the form $\Delta_1 = -\Delta_2 \neq 0$. This solution shifts the mean-field state away from the pattern $(\pi, \pi, \pi)$ for the flux state, and opens up a gap. 

The Chern number of each of the resulting bands can be computed by using the standard expression\cite{Thouless1982} in terms of the flux (through the Brillouin zone) of the Berry connections
\begin{equation}
  C = \frac{1}{2\pi} \int_{BZ} d^{2}k \mathcal{F}_{xy}(k)
\end{equation}
where $\mathcal{F}_{ij} = \partial_{i}\mathcal{A}_{j} - \partial_{j}\mathcal{A}_{j}$ is the flux of the Berry connection $\mathcal{A}_{i} = -i \bra{\psi}\partial_{k_{i}}\ket{\psi}$. Here $\ket{\psi}$ refers to the normalized
eigenvector of the corresponding band.

In the Ising regime the Chern numbers of the bands are
\beq
	\begin{matrix}
		C_1 = 0 &	& C_2 =0	&	&	C_3=0	\\
		C_4 = 0 &	& C_5 =0	&	&	C_6=0
	\end{matrix}
\eeq
This implies that in the Ising regime, the total Chern number for the filled bands is 0. This means that we are left with the original Chern-Simons term from the flux attachment transformation. In this regime, the fermions are essentially transmuted back to the original hard-core bosons (spins) that we began with and our analysis doesn't pick out any specific state.

\subsection{Mean-field theory with a non-vanishing chirality field, $h \neq 0$}
\label{sec:MF_chiral}

In this section, we will turn on the chirality term. Looking at the doubled unit cell in Fig. \ref{fig:MF_half_filling}, there are four corresponding chirality terms (within each magnetic unit cell) which can be written as
\beq
\chi_{ijk}(x) = \chi_{b1}(x)+\chi_{b2}(x)+\chi_{c1}(x)+\chi_{c2}(x)
\eeq

Now, we have to account for the additional contributions from $h$ in Eq.\eqref{eq:Jb}. Importantly, the added contribution to the gauge fields due to a non-zero value $\phi_{(a)}(x)$ in Eq.\eqref{eq:Jb} will give rise to additional fluxes and shift the state away from the $(\pi, \pi, \pi)$ flux state observed in the $XY$ regime section in Sec \ref{sec:XY_J}. Notice that, if we were stay in the $(\pi, \pi, \pi)$ flux state, this would imply that the average density $\langle n_{(a)}(x) \rangle = \frac{1}{2}$ at every site. In this situation the expectation value of the chirality operator automatically vanishes due to the relation $\langle S^{z}_{(a)} \rangle =\frac{1}{2}-\frac{1}{2\pi}\langle B_{(a)} \rangle $. In this situation the chirality term would never pick up an expectation value at the mean-field level and time reversal symmetry would remain unbroken. Hence, in a state with broken time reversal invariance the site densities cannot  all be exactly equal to $\frac{1}{2}$.

\begin{figure}
  \begin{center}
    \includegraphics[width=0.48\textwidth]{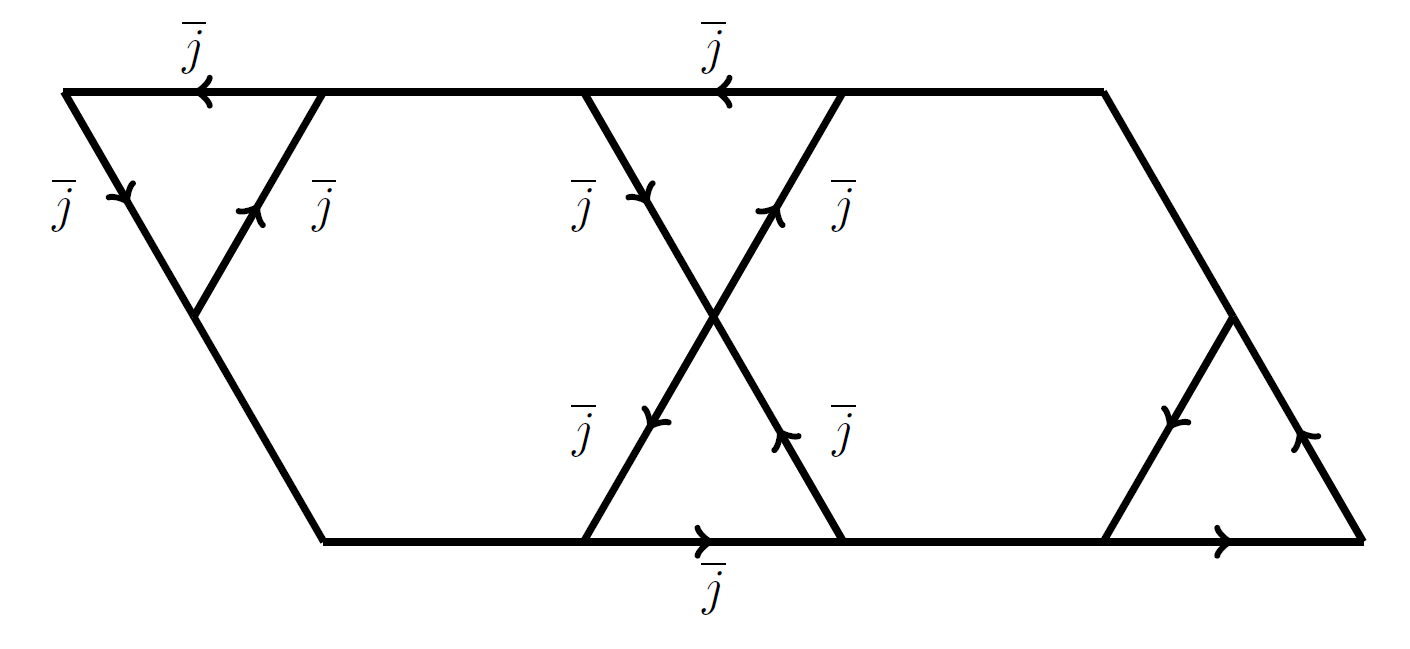}
  \end{center}
  \caption{Currents induced by the chirality term in each of the triangular plaquettes. The currents are indicated by $\jb$.}
  \label{fig:mag_unit_cell_currents}
\end{figure}
The fluxes in each of the plaquettes also gets modified due to the contribution from $\phi(x)$. The effective flux at each of the sublattice sites is now given as
\beq
\langle B_a \rangle =& \pi - 2\pi \Delta_1 - 2\pi \Delta_2 +2 \lb \phi_a + \phi_b + \phi_c \rb	\\
\langle B_b \rangle =& \pi + 2\pi \Delta_1 - \lb \phi_a + \phi_b + \phi_c \rb	\\
\langle B_c \rangle =& \pi + 2\pi \Delta_2 - \lb \phi_a + \phi_b + \phi_c \rb
\label{eq:flux_h}
\eeq
The above fluxes still ensure that we in the half-filled case.

In order to accommodate such a flux state, we also have to allow for non-zero currents in the mean-field state in Eq. \eqref{eq:MF_current}. As a result we will consider an ansatz with $\langle j_{k}(x,t) \rangle \neq 0$. The chirality terms in the Hamiltonian go across each of the triangular plaquettes in a counter-clockwise manner. Hence, we will choose an ansatz on each of the different links as seen in Fig. \ref{fig:mag_unit_cell_currents}. The mean-field equations for the current terms in Eq.\eqref{eq:MF_current} can now be satisfied by the below choice of gauge fields
\beq
	A_{0,a} =& 2 \lambda J(\Delta_1 + \Delta_2)+\jb	\\
	A_{0,b} =& -2 \lambda J\Delta_1-\jb	\\
	A_{0,c} =& -2 \lambda J\Delta_2-\jb
\eeq
In the above equations the effect of the chirality term directly enters in the form of a current. Now, we will proceed to look for mean-field phases that self-consistently satisfy the mean-field equations in Eq.\eqref{eq:MF_density} and Eq.\eqref{eq:MF_current} as well as constraints set by Eq.\eqref{eq:Jb} and Eq.\eqref{eq:flux_h}. Once again, we will analyze the cases of the $XY$ and Ising regimes separately.

\subsubsection{The $XY$ regime}
\label{sec:XY_chiral}

\begin{figure}[h]
\centering
  \includegraphics[width=\linewidth]{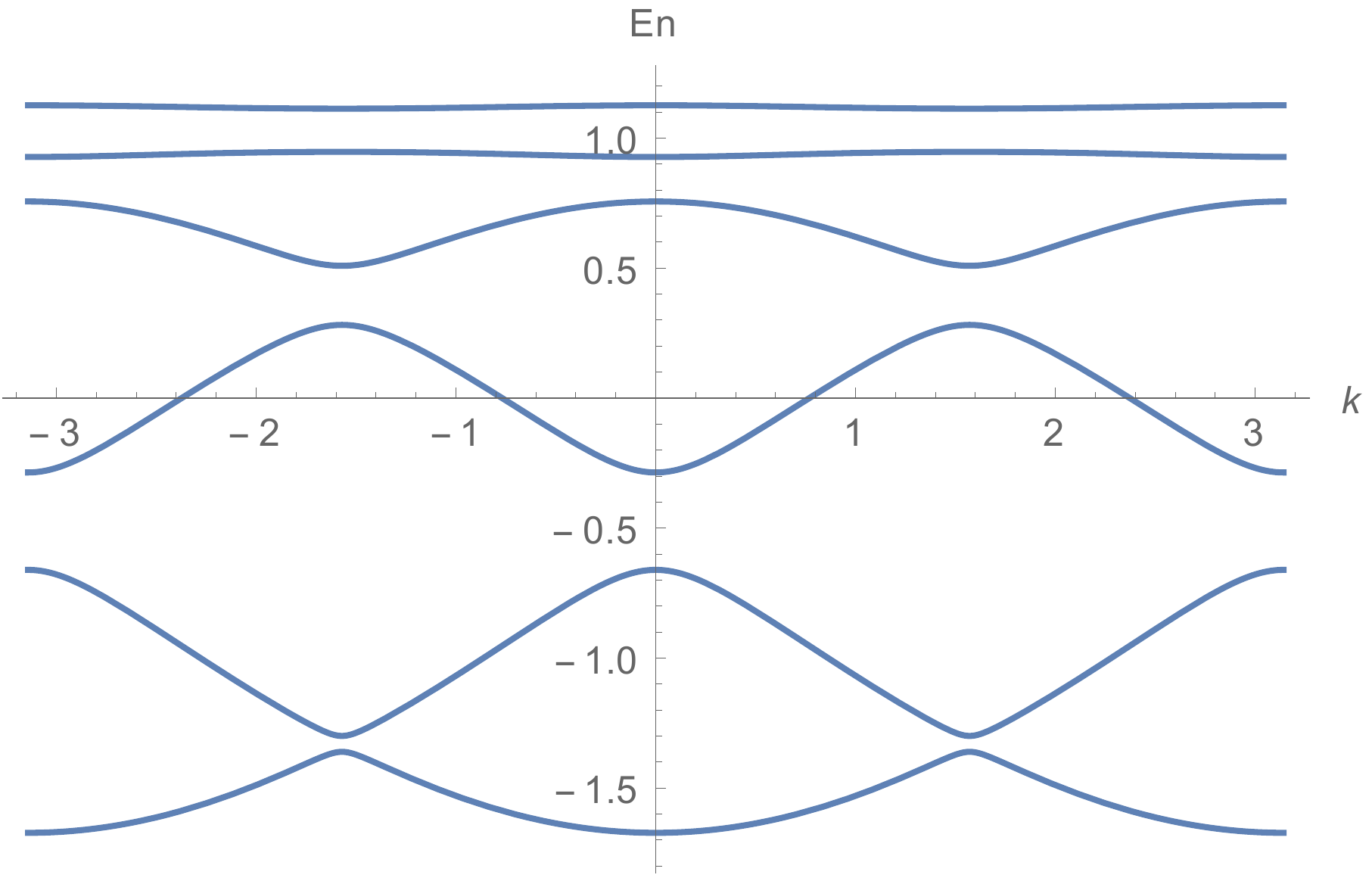}
\caption{(Color online) The chirality term   opens up a gap in the spectrum (see Fig \ref{fig:pi_flux_spectrum_Dirac}. The above plot is for $h=0.05J$. The plot is made along the $k_y = -\frac{k_x}{\sqrt{3}}$ line in the Brillouin zone.}
\label{fig:pi_flux_spectrum_h}
\end{figure}

In the $XY$ regime, $\frac{\lambda}{J} \leq 1$, we had the $(\pi, \pi, \pi)$ flux state which was gapless and had two Dirac points (see Fig. \ref{fig:pi_flux_spectrum_Dirac}). Here, we find that even for small values of $h$, there exist solutions with $\Delta_1 = \Delta_2 \neq 0$. This shifts the state away from the $(\pi, \pi, \pi)$ flux state and opens up a gap in the spectrum as shown in Fig. \ref{fig:pi_flux_spectrum_h}.

\begin{table*}
\setlength\extrarowheight{4pt}
\begin{ruledtabular}
 \begin{tabular}{ccccc}
  $\frac{h}{J}$ & $\langle n_b \rangle = \langle n_c \rangle$ & $\Delta_1 = \Delta_2$ & $E_G$ & $\langle \chi \rangle$\\
  \specialrule{1.1pt}{1pt}{1pt}
  0	&	0.500	&	0	&	0	&	0	\\
  0.05	&	0.460	&	-0.040	&	0.2286$J$	&	0.000782\\
  0.1	&	0.385	&	-0.115	&	0.5518$J$	&	0.001064\\	
  0.5	&	0.300	&	-0.200	&	0.7351$J	$	&	0.002149\\
  1		&	0.275	&	-0.225	&	0.7638$J$	&	0.002642\\
 \end{tabular}
 \end{ruledtabular}
 \caption{Approximate values for the mean-field parameters for different values of chirality ($h$) for $\frac{\lambda}{J}=1$. Here $E_G$ denotes   the energy gap in units of $J$, and $\langle \chi \rangle$ is the expectation value of the chirality operator. As the chirality term gets  stronger, the average density on each of the triangular plaquettes approaches $0.25$, and the density on the hexagonal plaquettes  approaches 1.}
 \label{table:MF_h}
\end{table*}

The values of the mean-field parameters for a few different values of the field $h$ (the strength of the chirality breaking term) are shown in Table \ref{table:MF_h}.  A plot of the mean-field spectrum for the specific case of $h=0.05J$ is shown in Fig. \ref{fig:pi_flux_spectrum_h}. As the value of $h$ is increased from $0$, the average flux on each of the triangular plaquettes decreases from $\pi \to \frac{\pi}{2}$. The corresponding flux in the hexagonal plaquettes goes from $\pi \to 2\pi$. In the limit of a strong chirality term, one would expect to get a state with flux of $2\pi$ in each hexagonal plaquette, and a flux of $\frac{\pi}{2}$ in each of the triangular plaquettes. We will refer to this as the $\lb 2\pi, \frac{\pi}{2}, \frac{\pi}{2} \rb$ flux phase. The values of the energy gap and the expectation values of the chirality operator are also shown for the different values of $h$ in Table \ref{table:MF_h}. (The energy gaps essentially measure the gaps between the Dirac points.)

The Chern numbers of the six bands with the chirality breaking field $h$ turned on are
\beq
	\begin{matrix}
		C_1 = +1 &	& C_2 =-1	&	&	C_3=+1	\\
		C_4 = +1 &	& C_5 =-1	&	&	C_6=-1
	\end{matrix}
\eeq

As we are still at half-filling, the bottom three bands must be filled,  leading to a total Chern number of the occupied bands of  $C_1 + C_2 + C_3 = +1$. This along with the original Chern-Simons term from the flux attachment transformation is expected to give rise to an effective Chern-Simons term with an effective parameter
\beq
	\theta_{\rm eff} 	&= \theta_F + \theta_{CS} = \frac{1}{2\pi} + \frac{1}{2\pi}
\eeq
A more detailed and rigorous computation of the above statement will be presented in a later section, Sec. \ref{sec:Continuum}, where we will include the effects of fluctuations and show that the resultant continuum action is indeed a Chern-Simons theory with the above effective parameter. This result shows that in the presence of the chirality term, we do obtain a chiral spin liquid. Such a state is equivalent to a Laughlin fractional quantum Hall state for bosons with a spin Hall conductivity $\sigma^s_{xy} = \frac{1}{2}$. The state obtained here has the same topological properties  as the state that we found\cite{Kumar2014} in the  magnetization plateau at $m=\frac{1}{3}$.

In the limit that we only have the chirality term, we have that $J = 0$. Now $\phi_{(a)} = \pm \frac{\pi}{2}$ in Eq.\eqref{eq:Jb}. We can now look for solutions such that
\beq
\phi_a = -\frac{\pi}{2} &\quad \quad \phi_b = \phi_c = \frac{\pi}{2}
\eeq
Here, we again recover the $(2\pi, \frac{\pi}{2}, \frac{\pi}{2})$ flux state as expected. The resulting chiral state is the same found by Bauer and coworkers  in Ref.[\onlinecite{Bauer-2014}].

\subsubsection{Ising regime}
\label{sec:Ising_chiral}

In the Ising regime, $\frac{\lambda}{J} > 1$, the Heisenberg model gave rise to a state that was  gapped and  a vanishing Chern number, as shown in Sec. \ref{sec:Ising_J}. Here a small chirality term would not affect the mean-field state as long as it is weak enough. In order to see the chiral spin liquid state obtained in the $XY$ regime in Sec. \ref{sec:XY_chiral}, one would need a strong enough chirality term to close the Ising anisotropy gaps and to open a chiral gap so as to give rises to states with non-trivial Chern numbers. Hence, the state here would be determined based on the competition between the anisotropy parameter $\lambda$ and the strength of the chirality parameter $h$.

\subsection{Combined effects of a chirality symmetry breaking term and an external magnetic field}
\label{sec:MF_chiral_hext}

So far, we have primarily focused on the case of half-filling and hence in the absence of an external magnetic field, $h_{\rm ext}=0$. Now we will briefly consider the scenario when the external magnetic field is present, $h_{\rm ext} \neq 0$, in Eq. \eqref{eq:Chiral_Hamiltonian} or, equivalently, that we are at fermionic fillings other than $\frac{1}{2}$ in the $XY$ limit. This will allow us to connect our recent results on a chiral spin liquid phase in a magnetization plateau with the chiral state arising  in the presence of a chirality symmetry breaking field. The mean field theory we discuss here has points of contact, including the role of Chern numbers,  with a classic paper by Haldane and Arovas.\cite{Haldane1995}

In the previous section we noted that the main effect of adding the chirality symmetry breaking term to the mean-field state was to shift the fluxes on each of the sub-lattices. We began with a $(\pi, \pi, \pi)$ flux phase for the Heisenberg model and it was modified to a $(2\pi, \frac{\pi}{2}, \frac{\pi}{2})$ flux phase in the presence of a strong chirality term. Essentially the chirality term shifted the fluxes from the triangles to the hexagons.
Using this analogy, we will now look for similar flux phases at other fillings. In the presence of a strong chirality term, we will consider flux phases where the flux in maximized in the hexagons and minimized in the triangles at different fillings.

\begin{figure}[ht]
\begin{center}
  \includegraphics[width=\linewidth]{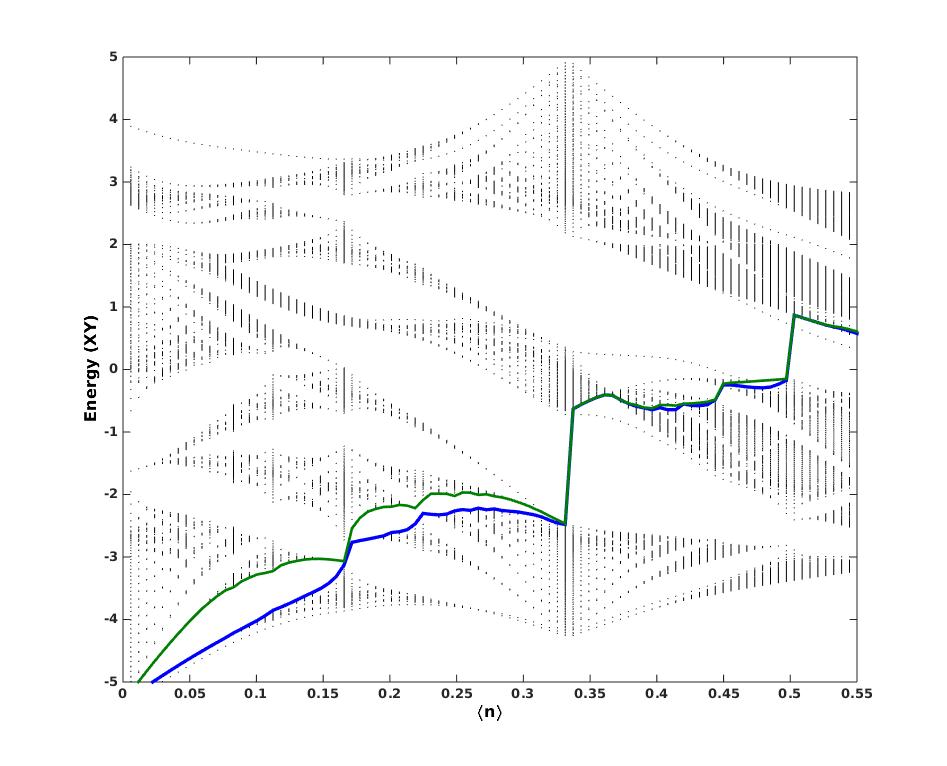}
  \end{center}
  \caption{(color online) Hofstadter spectrum for  $h = 5J$. The $x$-axis plots the mean-field fermion density $\langle n \rangle$ and the $y$-axis plots the $XY$ energies. The bottom solid line indicates the fermi level and the top solid line corresponds to the next excited state available. The vertical jumps in the figure correspond to possible magnetization plateaus.}
  \label{fig:hspectrum_chi_5}
\end{figure}

In the absence of the chirality term, we have a uniform flux phase with 
$\langle B_a \rangle= \langle B_b \rangle= \langle B_c \rangle= \phi = 2\pi \frac{p}{q}$ with $p,q \in \mathbb{Z}$. 
When, we turn on the chirality term, we expect the fluxes from the triangles to shift to the hexagons. Hence, we have 
\beq
\langle B_a \rangle=& \phi + 2\delta \\
\langle B_b \rangle=& \phi - \delta \\
\langle B_c \rangle=& \phi - \delta
\eeq
so that the total flux in each unit-cell is still the same. Such a flux state can be realized by the below choice of gauge fields
\beq
&\bA_1 (\vx) = 0, &  \bA_2(\vx) = \phi-\delta, \\
& \bA_3(\vx)=0, & \bA_4 (\vx) = 0, \\
& \bA_5(\vx) = -\phi+\delta+3\phi x_1, & \bA_6(\vx)=3\phi x_1
\eeq
with $\vx = (x_1, x_2)$. $x_1$ and $x_2$ are the coordinates along the $e_1$ and $e_2$ directions in Fig. \ref{fig:MF_half_filling} respectively.
 The fluxes on each plaquette range between $0$ and $2\pi$, which translates to having a site filling between $0$ and $1$. Hence, we set $\delta = \text{min}(\phi, \pi - \frac{\phi}{2})$. Using this choice, one can plot the Hofstadter spectrum in the limit of a strong chirality term. In Fig. \ref{fig:hspectrum_chi_5} we plot the Hofstadter spectrum for the case with $\frac{h}{J}=5$. The bottom solid line indicates the Fermi level (all the occupied states) and the top solid line indicates the next excited state available. 
 
 At most fillings the total Chern numbers of all the occupied bands is $-1$. This would lead to a Chern-Simons term with pre-factor $-\frac{1}{2\pi}$ and such a term would be expected to cancel when combined with the original Chern-Simons term from the flux attachment transformation, which also has a pre-factor $\frac{1}{2\pi}$. The exceptions are at the fillings $\langle n \rangle = \frac{1}{6}, \frac{1}{3}, \frac{4}{9}, \frac{1}{2}$, represented by vertical jumps in the solid lines in Fig. \ref{fig:hspectrum_chi_5}. At these fillings, the total Chern number of all the filled bands is different from $-1$ and lead to an effective Chern-Simons term. The resulting magnetization plateaus and their corresponding Chern-numbers are summarized in Table \ref{table:chispec_values}.

\begin{table}[hbt]
\setlength\extrarowheight{5pt}
\begin{ruledtabular}
\begin{tabular}{ccc}
$\langle n \rangle$ & Chern No. & $m$\\
\specialrule{1.2pt}{1pt}{1pt}
$\frac{1}{6}$ & +1 & $\frac{2}{3}$  \\
$\frac{1}{3}$ & +1 & $\frac{1}{3}$ \\
$\frac{4}{9}$ & +2 & $\frac{1}{9}$ \\
$\frac{1}{2}$ & +1 & $0$
\end{tabular}
\end{ruledtabular}
\caption{Magnetization plateaus obtained in Fig. \ref{fig:hspectrum_chi_5} and their Chern numbers. At these fillings the Chern-Simons terms do not  cancel out and the system  is in  a chiral spin liquid.}
\label{table:chispec_values}
\end{table}

The magnetization plateaus at filling fractions $\frac{1}{3}$ and $\frac{1}{6}$ have also been previously obtained in the absence of  the chirality symmetry breaking term, $h=0$ and $h_{\rm ext}\neq0$ in Ref. [\onlinecite{Kumar2014}]. It is apparent that these plateaus survive in the presence of the chirality symmetry breaking term. Additionally we observe two other plateaus at fillings $\frac{4}{9}$ and $\frac{1}{2}$. The plateau at $\frac{1}{2}$ filling is the same one that was observed in the previous sections for the case with no magnetic field (see Sec. \ref{sec:XY_chiral}). 

The plateau at $\frac{4}{9}$ filling has a magnetic unit cell with three basic unit cells. This gives rise to a total of nine bands of which four are filled. The Chern numbers of each of the nine bands in the mean-field state  are 
\beq
C_1 &= -1 \quad C_2 = +2 \quad C_3=+2 \\
C_4 &=-1 \quad C_5= -1 \quad C_6 = +2 \\
C_7 &=-1 \quad C_8=-1 \quad C_9=-1
\eeq
The Chern numbers of the four filled bands ($C_1$, $C_2$, $C_3$, and $C_4$) add up to $+2$. Again this result will combine with the Chern-Simons term from the flux attachment transformation leading to an effective Chern-Simons with an effective spin Hall conductance of $\sxy = \frac{2}{3}$. This result is also summarized in Table \ref{table:chispec_values}. In Ref.[\onlinecite{Kumar2014}] we identified this state as having the same topological properties as the first state in the Jain sequence of fractional quantum Hall states of bosons.

\subsection{Chiral Spin Liquids with Dzyaloshinski-Moriya Interactions}
\label{sec:MF_DM}

In this Section, we consider the effects of  a Dzyaloshinski-Moriya term (instead of the chirality term) on the nearest neighbor $XXZ$ Heisenberg Hamiltonian in Eq. \eqref{eq:Heisenberg_Hamiltonian}. The Dzyaloshinski-Moriya term is written as
\beq
H_{DM} = J_{DM} \sum_{i,j} \hat{z}\cdot({\bm S}_i \times {\bm S}_j)
\eeq
where the sum runs over nearest neighbors in each triangle in a clockwise manner. As an example the Dzyaloshinski-Moriya term in a triangle associated with site $b_1$ in Fig. \ref{fig:MF_half_filling} can be written as
\beq
H_{DM,b_1} = \frac{i}{2} J_{DM} \sum_{\vx} \bigg( & 
S^+_{a_2}(\vx)S_{b_1}^-(\vx) - S^+_{b_1}(\vx)S_{a_2}^-(\vx) \\
 + S^+_{c_2}(\vx)& S_{a_2}^-(\vx) - S^+_{a_2}(\vx)S_{c_2}^-(\vx) \\
 + S^+_{b_1}(\vx)& S_{c_2}^-(\vx) - S^+_{b_1}(\vx)S_{c_2}^-(\vx)
\bigg)
\label{eq:DM}
\eeq
Clearly this term breaks time reversal so we expect that we may be able to find chiral phases.

From the form of Eq.\eqref{eq:DM}, we can now readily apply the flux attachment transformation just like we had for the case of the chirality term.
As a result the parameters in Eq.\eqref{eq:Jb} now get modified as
\beq
\Jb_{(a)}(x) =& \frac{1}{2}\sqrt{ J^2 + \left[ h \lb \frac{1}{2} - n_{(a)}(x) \rb + J_{DM} \right]^2 }	\\
\phi_{(a)}(x) =& \text{tan} \inv \lb \frac{h}{J}\lb \frac{1}{2} - n_{(a)}(x) \rb + \frac{J_{DM}}{J}\rb
\label{eq:Jb_DM}
\eeq
In this section, we will set the chirality symmetry breaking term to zero, $h=0$.

Two separate regimes have to be considered.

\subsubsection{$J_{DM} \lesssim 1.7J$}

\begin{figure}[h]
  \centering
  \includegraphics[width=\linewidth]{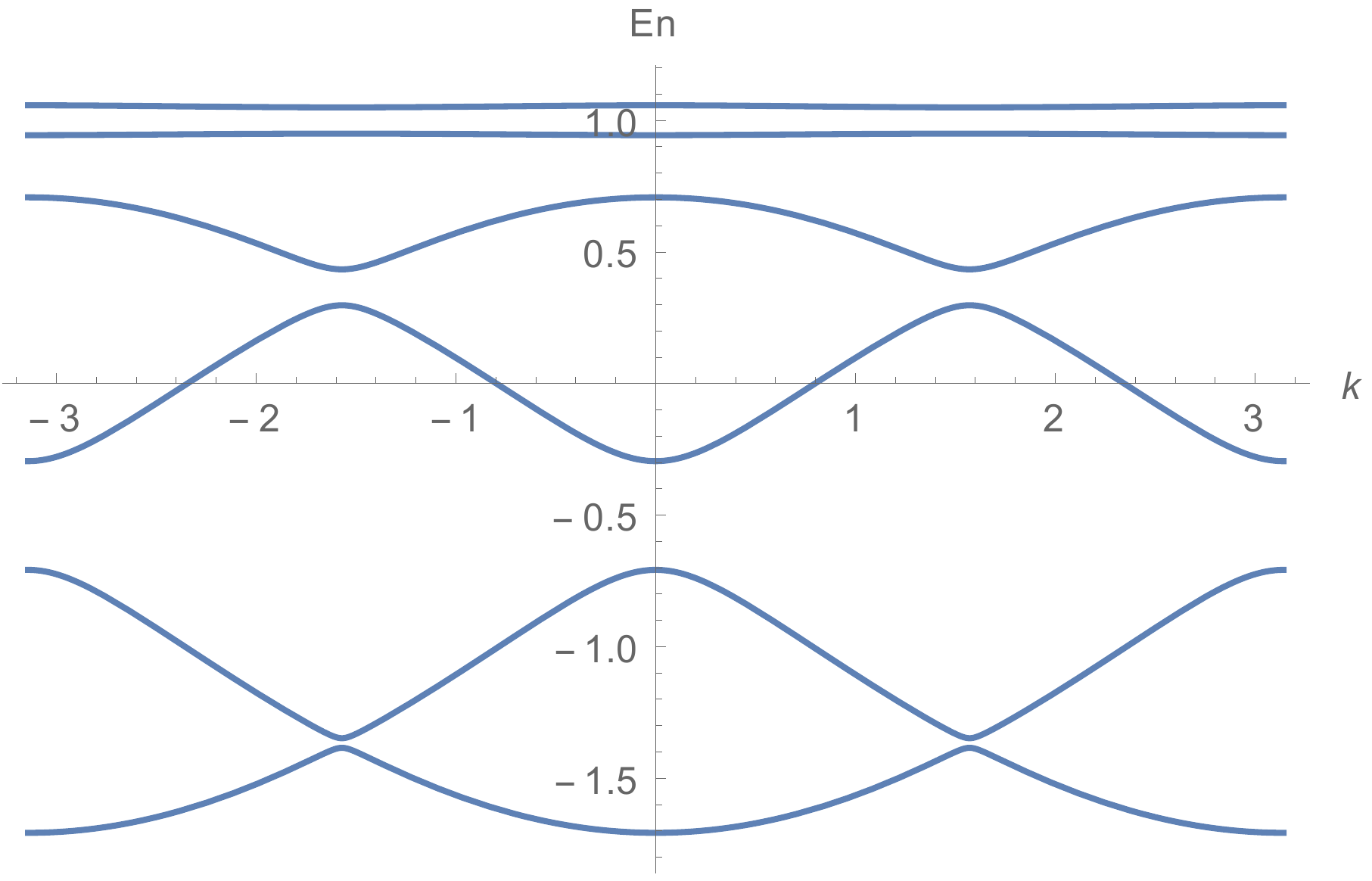}
  \caption{(color online) Spectrum with $J_{DM}=0.05J$. The Dzyaloshinski-Moriya term opens up a gap in the spectrum of the Heisenberg Hamiltonian (see Fig. \ref{fig:pi_flux_spectrum_Dirac}).}
\label{fig:spectrum_jdm}
\end{figure}

Recall that in the $XY$ regime the Heisenberg model gave rise to the $(\pi, \pi, \pi)$ flux state which is gapless and has Dirac points (Fig. \ref{fig:pi_flux_spectrum_Dirac}). Treating the Dzyaloshinski-Moriya term as a perturbation, we find that this term also opens up a gap in the $(\pi, \pi, \pi)$ flux state as can be seen in Fig. \ref{fig:spectrum_jdm}. But a the resultant state obtained still has a flux of $\pi$ in each of the plaquettes. For the situation shown in Fig \ref{fig:spectrum_jdm}, the energy gap is 0.1366J. This is an important difference between the effects of adding the chiral term and the Dzayloshinkii-Moriya terms, since the chirality term shifted the fluxes on each plaquette away from $\pi$ whereas the Dzyaloshinski-Moriya term does not.

The Chern numbers of the six bands in the presence of a small $J_{DM}$ term are
\beq
	\begin{matrix}
		C_1 = +1 &	& C_2 =-1	&	C_3=+1	\\
		C_4 = +1 &	& C_5 =-1	&	C_6=-1
	\end{matrix}
\eeq
Once again, we find that the total Chern number of all the filled bands is $C_1+C_2+C_3 = +1$. This would again lead to a fractional quantum Hall type phase with $\sigma^s_{xy} = \frac{1}{2}$, just as we had observed in the case with the chirality term. 

\subsection{$J_{DM} \gtrsim 1.7J$ }

For larger values of the Dzyaloshinski-Moriya parameter, namely for $\frac{J_{DM}}{J} \gtrsim 1.7$, the Chern numbers of the bands again get rearranged and the chiral phase no longer survives, as shown below
\beq
	\begin{matrix}
		C_1 = -1 &	& C_2 =+1	&	C_3=-1	\\
		C_4 = -1 &	& C_5 =+1	&	C_6=+1
	\end{matrix}
\eeq

In the limit that we only have the Dzyaloshinski-Moriya term, i.e. $J=0$, the values of all $\phi_{(a)} = \frac{\pi}{2}$ in Eq.\eqref{eq:Jb_DM}. In this case the values of the mean-field parameters that satisfy the consistency equations, Eq.\eqref{eq:MF_density} and Eq. \eqref{eq:MF_current}, are $
\Delta_1 = \Delta_2 = -\frac{1}{4}$. Hence, in the presence of only the Dzayloshinski-Moriya term, we again end up in the $(\pi, \pi, \pi)$ flux state that was observed in the $XY$ regime of the Heisenberg model in Sec \ref{sec:XY_J}. 

\subsection{Dzyaloshinski-Moriya term with an uniform magnetic field, $h_{\rm ext}\neq0$}
\label{sec:MF_DM_hext}

Finally, we will also consider the effects of the Dzyaloshinski-Moriya term in the presence of an uniform external magnetic field $h_{\rm ext}\neq0$,  just as we had done for the chirality terms in Sec. \ref{sec:MF_chiral_hext}. We will once again focus on the $XY$ limit where the mean-field equations are simpler due to the absence of the interaction term, i.e. $\lambda=0$. We will look for states that are uniform, time-independent and don't have any currents.

This scenario is very similar to the case of the integer quantum Hall effect with non-interacting fermions in the presence of a (statistical) gauge field. This approach was also used by Misguich et. al. in their studies on the triangular lattice.\cite{Misguich2001} More recently, we carried out a similar analysis  on the kagome lattice with an $XY$ nearest neighbor Heisenberg model.\cite{Kumar2014} Here, we will perform the same analysis, but with the Dzyaloshinski-Moriya term added to the $XY$ nearest neighbor Heisenberg model.

\begin{figure}
\begin{center}
  \includegraphics[width=\linewidth]{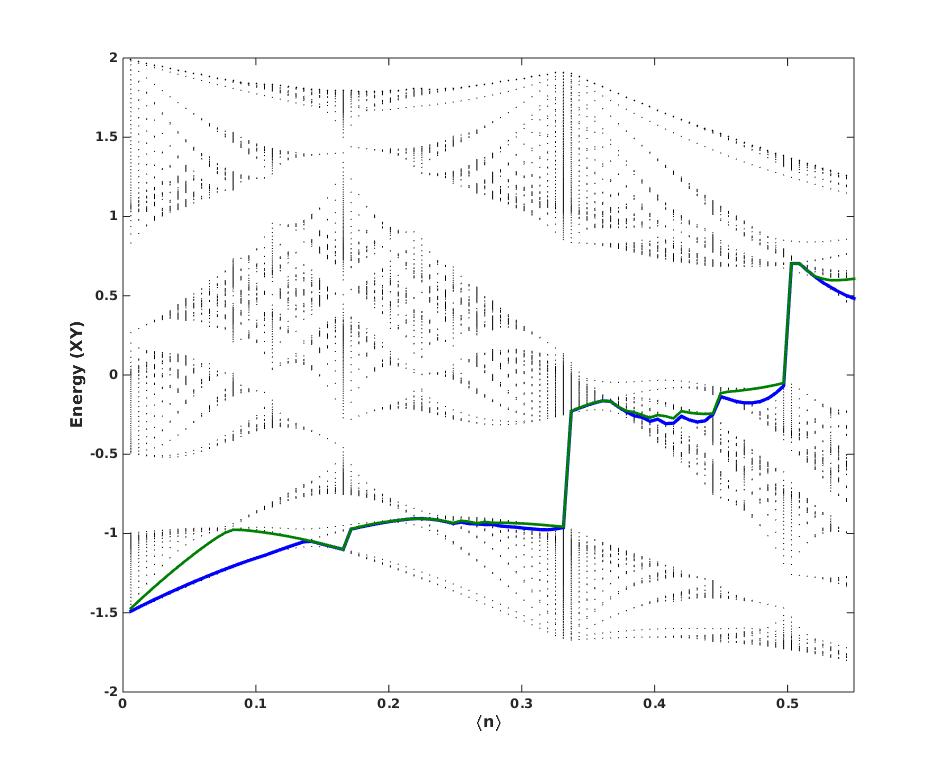}
  \end{center}
  \caption{(Color online) Hosftadter spectrum in the $XY$ limit for the case $J_{DM}=0.3J$. (The x-axis plots the mean-field fermion density $\langle n \rangle$ and the y-axis plots the energies of the $XY$ model.) The bottom solid line represents the Fermi level (all the filled bands) and the top solid line represents the next excited energy state available. The plateaus correspond to jumps in the solid line. We see two additional plateaus at densities $\langle n \rangle = \frac{4}{9}, \frac{1}{2}$ for certain values of $J_{DM}$. }
\label{fig:hspectrum_jdm}
\end{figure}

Once again, we find a few different plateaus as can be seen in the Hofstadter spectrum in Fig \ref{fig:hspectrum_jdm} for $J_{DM}=0.3 J$. The vertical lines in the figure correspond to the magnetization plateaus. The range of $J_{DM}$ values for which we observe the above plateaus is shown in Table \ref{table:hspec_values}. The table also lists the total Chern numbers of all the filled bands at each of the plateaus as well as the corresponding magnetization.

\begin{table}
\setlength\extrarowheight{5pt}
\begin{ruledtabular}
\begin{tabular}{cccc}
$\langle n \rangle$ & Range of values (in $J$) & Chern No. & $m$ \\
\specialrule{1.2pt}{1pt}{1pt}
$\frac{1}{6}$ & $0 \leq J_{DM} \lesssim 0.35$ & +1 & $\frac{2}{3}$	\\
$\frac{2}{9}$ & $0 \leq J_{DM} \lesssim 0.3$ & +2 &	$\frac{5}{9}$	\\
$\frac{1}{3}$ & $0 \leq J_{DM} \lesssim 0.8$ & +1 &	$\frac{1}{3}$	\\
$\frac{4}{9}$ & $0.05 \lesssim J_{DM} \lesssim 0.6$ & +2 & $\frac{1}{9}$	\\
$\frac{1}{2}$ & $0 < J_{DM} \lesssim 1.7$ & +1 & 0	
\end{tabular}
\end{ruledtabular}
\caption{Approximate values of $J_{DM}$ for which we observe the plateaus at the mean-field level. This table also lists the corresponding Chern numbers and their magnetizations, $m$. }
\label{table:hspec_values}
\end{table}

This concludes our mean-field analysis into the various possible magnetization plateaus. We will now proceed to consider the effects of fluctuations on the mean-field state when a small chirality term was added to the Heisenberg model in the $XY$ limit. This was the situation discussed in Sec \ref{sec:MF_chiral}. For the rest of the paper, we will not consider the Dzyaloshinski-Moriya term or the external magnetic field term again.

\section{Effective Field Theory}
\label{sec:Continuum}

In this section we return to the case of the nearest neighbor Heisenberg model in the presence of a small chirality term. In Sec. \ref{sec:MF_chiral}, it was shown that the addition of the chirality term opened up a gap in the mean-field spectrum and lead to a state with non-trivial Chern number. We will now expand the fermionic action around this mean-field state and consider its continuum limit. This process will allow us to go beyond the mean-field level and consider the fluctuation effects of the statistical gauge fields. The analysis presented here is analogous to the one presented in our earlier work.\cite{Kumar2014}  As a result we will only write down the relevant expressions for the current scenario.

In Sec. \ref{sec:XY_chiral}, we found that in the absence of the chirality term the spectrum was gapless with two Dirac points and that the addition of the chirality term opened up a gap at the Dirac points. 
These two Dirac points in the mean-field phase were located at the momenta ${\bf K} = \pm \left(\frac{\pi}{2}, -\frac{\pi}{2\sqrt{3}} \right)$. The fermionic degrees of freedom on each site can be expanded around each of the two Dirac points using the following expansions on each sub-lattice
\beq
\bmb
\psi_{a1} \\ \psi_{b1} \\ \psi_{c1} \\ \psi_{a2} \\ \psi_{b2} \\ \psi_{c2}
\emb
& \sim \frac{a_0}{\sqrt{6}}\bmb
- e^{i \frac{5\pi}{12}} & i \\
\sqrt{2} & 0 \\
- e^{i \frac{5\pi}{12}} & - e^{-i \frac{\pi}{6}}  \\
- e^{i \frac{5\pi}{12}} & i \\
0 & -\sqrt{2}e^{-i \frac{\pi}{12}} \\
- e^{-i \frac{\pi}{12}} & e^{-i \frac{\pi}{3}}  
\emb
\bmb
\Psi_1^1 \\ \Psi_1^2
\emb
\\
\bmb
\psi_{a1} \\ \psi_{b1} \\ \psi_{c1} \\ \psi_{a2} \\ \psi_{b2} \\ \psi_{c2}
\emb
& \sim \frac{a_0}{\sqrt{6}}\bmb
- e^{i \frac{5\pi}{12}} & -i \\
0 & -\sqrt{2}e^{-i \frac{\pi}{12}} \\
e^{i \frac{\pi}{12}} & - e^{i\frac{\pi}{3}}  \\
e^{i \frac{5\pi}{12}} & -i \\
-\sqrt{2} & 0 \\
- e^{-i \frac{5\pi}{12}} & -e^{-i \frac{\pi}{6}}  
\emb
\bmb
\Psi_2^1 \\ \Psi_2^2
\emb
\label{eq:Dirac_pts}
\eeq
where in $\Psi_r^{\alpha}$, $r$ refers to the Dirac species index and the label $\alpha$ refers to the spinor index within each species. $\psi_{a1}, \psi_{b1}, \psi_{c1}, \psi_{a2}, \psi_{b2}, \psi_{c2}$ refer to the original fermionic fields on the different sub-lattices sites in the mean-field state at half-filling as shown in Fig \ref{fig:MF_half_filling}.

Now we will include the fluctuating components i.e. we will expand the statistical gauge fields as follows $\Amu = \langle \Amu \rangle + \delta \Amu$. The mean-field values of $\langle \Amu \rangle$ are the same as those given in Sec. \ref{sec:MF_chiral}. From now on, we will primarily focus on the fluctuating components. In order to simplify the notation, we will drop the $\delta$ label in the fluctuating components i.e. all the gauge fields presented beyond this point are purely the fluctuating components.

\subsection{Spatial fluctuating components}

First, we will begin by looking at just the spatial fluctuating components. Furthermore, we will also expand all the spatial fluctuating components in the magnetic unit cell in Fig. \ref{fig:MF_half_filling} in terms of slow and fast components. This will allow us to treat the slow components as the more relevant fields.

The fields along the $e_1$ direction (in Fig. \ref{fig:MF_half_filling}) can be expanded as
{\small
\beq
 A^1_{1} =& \frac{a_0}{2} \left( A_x + A^{f1}_1 + A^{f2}_1 - A^{f3}_1 \right)	\\
 A^1_{4} =& \frac{a_0}{2} \left( A_x + A^{f1}_1 - A^{f2}_1 + A^{f3}_1 \right) \\
 A^2_{1} =& \frac{a_0}{2} \left( A_x - A^{f1}_1 + A^{f2}_1 - A^{f3}_1 \right)
 \\
 A^2_{4} =& \frac{a_0}{2} \left( A_x - A^{f1}_1 - A^{f2}_1 + A^{f3}_1 \right) 
\eeq
}
Similarly, the fields along the $e_2$ direction can be written as
{\small
\beq
 A^1_{2} =& \frac{a_0}{2} \left(-\frac{1}{2} A_x + \frac{\sqrt{3}}{2}A_y - A^{f1}_2 - A^{f2}_2 + A^{f3}_2 \right) 
 \\
 A^1_{5} =& \frac{a_0}{2} \left(-\frac{1}{2} A_x + \frac{\sqrt{3}}{2}A_y + A^{f1}_2 + A^{f2}_2 - A^{f3}_2 \right) 
 \\
 A^2_{2} =& \frac{a_0}{2} \left(-\frac{1}{2} A_x + \frac{\sqrt{3}}{2}A_y + A^{f1}_2 - A^{f2}_2 + A^{f3}_2 \right)
 \\
  A^2_{5} =& \frac{a_0}{2} \left(-\frac{1}{2} A_x + \frac{\sqrt{3}}{2}A_y - A^{f1}_2 + A^{f2}_2 - A^{f3}_2 \right) 
\eeq
}

Finally, the fields along $e_1+e_2$ directions can be expressed as
{\small
\beq
 A^1_{3} =& \frac{a_0}{2} \left(\frac{1}{2} A_x + \frac{\sqrt{3}}{2}A_y+\frac{3}{2}A^f - A^{f1}_3 - A^{f2}_3 + A^{f3}_3 \right)
 \\
 A^1_{6} =& \frac{a_0}{2} \left(\frac{1}{2} A_x + \frac{\sqrt{3}}{2}A_y+\frac{3}{2}A^f + A^{f1}_3 + A^{f2}_3 - A^{f3}_3 \right)
 \\
 A^2_{3} =& \frac{a_0}{2} \left(\frac{1}{2} A_x + \frac{\sqrt{3}}{2}A_y+\frac{3}{2}A^f - A^{f1}_3 - A^{f2}_3 + A^{f3}_3 \right)
 \\
 A^2_{6} =& \frac{a_0}{2} \left(\frac{1}{2} A_x + \frac{\sqrt{3}}{2}A_y+\frac{3}{2} A^f - A^{f1}_3 + A^{f2}_3 - A^{f3}_3 \right)
 \label{eq:slow_fast_space}
\eeq
}

In the above expressions $A^{(A)}_i$ refer to the fluctuating components along the different links of the unit cell in the mean-field state ($1$ and $2$ refer to the two unit cells in the magnetic unit cel shown in Fig. \ref{fig:MF_half_filling}). The slow components are represented by $A_x$ and $A_y$ and $A^f$, $A^{f1}_i$, $A^{f2}_i$ and $A^{f3}_i$ are the fast fields along the different spatial directions. 

\subsection{Temporal fluctuating components}
Similarly, the fluctuating time components can also be expanded in terms of slow and fast fields as follows
{\small
\beq
A_{0,a1} =& a_0 \left( A_0 + \Azfzo + \Azfzo + 3 \sqrt{2} (\Azfo - \Azft) \right) \\
A_{0,b1} =& a_0 \left( A_0 - \Azfoz - 3 \Azfth \right) \\
A_{0,c1} =& a_0 \left( A_0 - \Azfzo + 3 \sqrt{2 - \sqrt{3}} \Azfo - 3\sqrt{2 + \sqrt{3}} \Azft \right) \\
A_{0,a2} =& a_0 \left( A_0 + \Azfzo + \Azfzo - 3 \sqrt{2} (\Azfo - \Azft) \right) \\
A_{0,b2} =& a_0 \left( A_0 - \Azfoz + 3 \Azfth \right) \\
A_{0,c1} =& a_0 \left( A_0 - \Azfzo - 3 \sqrt{2 - \sqrt{3}} \Azfo + 3\sqrt{2 + \sqrt{3}} \Azft \right)
\label{eq:slow_fast_time}
\eeq
}
where $a1$, $b1$, $c1$, $a2$, $b2$ and $c2$ again refer to the different sub-lattice indices in the mean-field state in Fig \ref{fig:MF_half_filling}. The only temporal slow component is $A_0$. All the other fields with super-script $f$ refer to the fast fields. The pre-factors and constants in Eq.\eqref{eq:slow_fast_time} are chosen to make the notation and computation below easier.

Using Eq.\eqref{eq:Dirac_pts}, the mean-field action with the choice of the mean-field gauge fields in Sec \ref{sec:MF_chiral} in the continuum limit becomes
\beq
S_{F,MF} = \int d^3x  \Psib_r (i \spartial - m) \Psi_r
\label{eq:MF}
\eeq
where $\Psib = \Psi^* \gamma_0$ with $\Psi = \bmb \Psi_1^1, & \Psi_1^2, & \Psi_1^2, & \Psi_2^2 \emb^T$. We are using the slash notation $\spartial = \gamma^{\mu}\partial_{\mu}=  \gammaz \partial_0 - \gamma_i \partial_i$ with the Minkowski metric $g_{\mu\nu}$. The gamma matrices act on the upper or spinor index ($\alpha$) in $\Psi_r^{\alpha}$ and are given by
\beq
\begin{matrix}
 \gammaz = \sigma_3 & & \gammao = i \sigma_2 & & \gammat = i \sigma_1
\end{matrix}
\label{eq:gamma}
\eeq
Importantly the mass terms $m$ are the same for both the Dirac points and are given as
\beq
  m = \lim_{a_0 \to 0} \left[ -\frac{\pi \Delta}{9 a_0}(3 + \sqrt{3}) \right] = -\frac{0.5258}{a_0}\Delta > 0
  \label{eq:mass}
\eeq
Hence, the masses $m$  are positive for {\em both}  Dirac points as the value of $\Delta<0$ from the mean-field analysis (as shown in Table \ref{table:MF_h}).

The resulting action for the spatial fluctuating components becomes
\beq
\delta S^{\rm slow}_{i} =& \int d^3x \lcb A_x \Psib \gamma_1 \Psi + A_y \Psib \gamma_2 \Psi  \rcb \\
\delta S^{\rm fast}_{i} =& \int d^3x \bigg\{ \frac{1}{2} A^f \Psib \gamma_1 \Psi + \frac{\sqrt{3}}{2}A^f \Psib \gamma_2 \Psi \\
	& - (\Aoft - \Aofth + \Atft - \Atfth - \Athft + \Athfth) \Psib \Psi \bigg\}
\label{eq:spatial_action}
\eeq
where we have absorbed some of the constant factors into the definitions of the fast fields to make the notation more convenient and the definitions of the gamma matrices are the same as in Eq. \eqref{eq:gamma}.

The resulting continuum action for the slow and fast fields become
\beq
\delta S_0^{\rm slow} =& - \int d^3x A_0 \left( \Psib \gamma_0 \Psi \right) \\
\delta S_0^{\rm fast} =& \int d^3x \bigg\{ \Azfo \left( \Psib \gamma_1 T^3 \Psi \right) + \Azft \left( \Psib \gamma_2 T^3 \Psi \right) \\
	& \quad + \Azfth \left( \Psib \gamma_0 T^3 \Psi \right) \bigg\}
\label{eq:Time_action}
\eeq 
where $T^3$ is the regular $\sigma_3$ Pauli matrix but  acting on the species index $r$ in $\Psi_r$.
Combining equations Eq.\eqref{eq:MF}, Eq.\eqref{eq:spatial_action} and Eq. \eqref{eq:Time_action}, the total continuum fermionic action for the slow components becomes
\beq
S_{F,{\rm slow}} = \int d^3x \Psib \left[i \gammamu \Dmu - m \right]\Psi
\label{eq:slow_action}
\eeq
where $\Dmu = \partial_{\mu} + i \Amu$ is the covariant derivative.
The fast components can be expressed as 
\beq
  S_{F,{\rm fast}} = \int d^3x & \bigg\{ \frac{1}{2} A^f \Psib \gamma_1 \Psi + A^f \frac{\sqrt{3}}{2 } \Psib \gamma_2 \Psi \\
  &+ \Azfo \Psib \gamma_1 T^3 \Psi +  \Azft \Psib \gamma_2 T^3 \Psi	\\
  &+ \Azfth \Psib \gamma_0 T^3 \Psi - \phi_i \Psib \Psi \bigg\}
\label{eq:fast_action} 
\eeq
where
\beq
  \phi_i =& \frac{\sqrt{2 + \sqrt{3}} }{3}\left( \Aft_1  - \Afth_1 + \Aft_2 - \Afth_2 - \Aft_3 + \Afth_3 \right)
\label{eq:phii}
\eeq
\\
Eq.\eqref{eq:slow_action} and Eq. \eqref{eq:fast_action} can also be expressed in momentum space as
\beq
S_F = \int \frac{d^3p}{(2\pi)^3} \Psib M \Psi
\label{eq:action_p}
\eeq
with $M = \Mb + \delta M$. 

The mean-field part $\Mb$ is given as
\beq
\Mb = 
\bmb
\ps - m & 0 \\
0 & \ps - m
\emb
\label{eq:MF_cont_prop}
\eeq
and the fluctuation part $\delta M$ is given as

\begin{widetext}
{\small
\beq
\delta M = \bmb
-\sA - \phi_i + \left(\frac{1}{2}A^f+\Azfo \right) \gamma_1 + \left(\frac{\sqrt{3}}{2} A^f + \Azft \right) \gamma_2 + \Azfth \gamma_0 & 0 \\
0 & -\sA - \phi_i +\left(\frac{1}{2}A^f - \Azfo \right) \gamma_1 +\left(\frac{\sqrt{3}}{2}A^f- \Azft \right) \gamma_2 - \Azfth \gamma_0  
\emb
\label{eq:Cont_prop}
\eeq
}
\end{widetext}

where $\phi_i$ is given in Eq.\eqref{eq:phii}.

The action in Eq.\eqref{eq:action_p} is quadratic in fermionic fields and fermions can be integrated out to give an effective action in terms of just the fluctuating gauge fields. The resulting effective action becomes
\beq
  S_{\rm eff} &= -i \text{Tr} \ln M 
\eeq
where $M$ is defined in Eq.\eqref{eq:Cont_prop}. Now we can expand $M$ in terms of the mean-field part and the fluctuating parts as shown in Eq.\eqref{eq:MF_cont_prop} and Eq.\eqref{eq:Cont_prop}. 

\beq
 S_{\rm eff} 	&= -i \text{Tr} \left\{ \ln \left(\Mb + \delta M \right) \right\} \\
 			&= -i \text{Tr} \left\{\ln \Mb \right\} -i \text{Tr} \left\{\ln \left( 1 + \Mb^{-1} \delta M \right) \right\}  
			\label{eq:expansion-trlog}
\eeq
Expanding this action up to second order in the fluctuating components gives
\beq
 S_{\rm eff} =& \frac{i}{2} \text{Tr}\left(\Mb^{-1}\delta M \Mb^{-1} \delta M \right) \\ 
      =& \frac{i}{2} \int_{p,q} \text{tr}\left(\Sb(p) \delta M(q) \Sb(p+q) \delta M(-q) \right)
      \label{eq:S_eff}
\eeq
where the lower-cased `tr' is a matrix trace, and $\Sb(p) = \Mb(p)^{-1}$ is the continuum mean-field propagator presented in Eq. \eqref{eq:MF_cont_prop}, and it is given by
\beq
\Sb(p) = \frac{1}{p^2 - m^2}
\bmb
\ps +m & 0 \\
0 & \ps + m
\emb
\eeq
In the expansion of Eq.\eqref{eq:expansion-trlog}  we will only keep the most relevant (mass) terms (without derivatives) for the fast components.

Similarly, one can also express the lattice version of the Chen-Simons term and the interaction terms using the slow and fast fluctuating components. Combining all of the above terms, one can obtain the final continuum action. All the massive fields can safely be integrated out. This leaves us with just the Chern-Simons and Maxwell terms. The computation of this Feynam diagram is standard and it is done in many places in the literature.\cite{Redlich-1984,Fradkin2013}

To lowest order, after integrating out all the massive fields, the most relevant term is the effective Chern-Simons term $S_{\rm eff}^{CS}$, since it has the smallest number of derivatives, and is given by
\beq
S^{CS}_{\rm eff} 	=& \lb \frac{\theta}{2} + \frac{\theta_F}{2} \rb \int d^3x \emnl A_{\mu} \partial_{\nu} A_{\lambda}  
\eeq
where $\theta= \frac{1}{2\pi}$ from the original flux attachment transformation and $\theta_F$ is the obtained from integrating out the fermions and is given as
\beq
\theta_F = \frac{1}{4\pi} (\text{sgn}(m) + \text{sgn}(m) ) = \frac{1}{2\pi}
\eeq
as $\text{sgn}(m) = +1$ ($m>0$ as shown in Eq.\eqref{eq:mass}). 

Hence, the Chern-Simons terms add up, and we get a state with spin Hall conductivity $\sigma^s_{xy} = \frac{1}{2}$. This state is equivalent to a bosonic Laughlin fractional quantum Hall state. This agrees and verifies our expected result obtained in Sec. \ref{sec:MF_chiral}.

The Maxwell terms can be conveniently expressed in terms of the electric $\bE$ and magnetic $B$ fields as follows 
\beq
S_{EM} = \int d^3x \lb \frac{1}{2}\epsilon \bE^2 - \frac{1}{2} \chi B^2 \rb
\eeq
where $\epsilon = \frac{1}{16\pi \sqrt{m^2}}$ and $\chi = \lb 24\sqrt{3} a_0 -\frac{1}{16\pi \sqrt{m^2}} \rb$.

The computation in this section confirms our expectation and analysis used to determine the nature of the chiral spin liquid states using the mean-field theory approaches in Sec. \ref{sec:MFT}.

\section{Spontaneous  breaking of time reversal invariance}
\label{sec:SSB}

In the cases discussed so far in this paper, we began with a  $(\pi,\pi,\pi)$ flux state which , at the level of the mean field theory, has massless Dirac fermions, and showed that breaking the time-reversal symmetry explicitly, by adding either a chirality term (Sec. \ref{sec:MF_chiral}) or a Dzyaloshinski-Moriya term (Sec. \ref{sec:MF_DM}), led to a gapped state. We the showed, that quantum corrections led directly to a chiral spin liquid with broken time-reversal symmetry for arbitrarily small values of the chiral field $h$ or the Dzyaloshinski-Moriya  interaction $J_{DM}$. The existence of an explicit gap in the spectrum of the fermions was essential to this analysis. Furthermore, after the leading quantum corrections are taken into account, we found that the naive Dirac fermions of the mean-field theory became anyons (semions in the cases that were discussed in detail). This line of reasoning parallels the theory of the fractional quantum Hall effect where, at the mean field level, one begins with composite fermions fulling up effective Landau levels,\cite{Jain1989} which turn into anyons by virtue of the quantum corrections.\cite{Lopez1991,Fradkin2013}

We now turn to the question of whether it is possible to obtain a chiral spin liquid by spontaneous time-reversal symmetry breaking. This concept was formulated originally by Wen, Wilczek and Zee\cite{Wen-1989} in the context of the $J_1-J_2$ Heisenberg model on the square lattice, where a chirally-invariant $\mathbb{Z}_2$ spin liquid appears to be favored instead.\cite{Jiang2011,Jiang2012} In this section we will show that ring-exchange processes on the bow ties (i.e. two triangles  sharing the same spin) of the kagome lattice may favor the spontaneous formation of the chiral spin liquid if the associated coupling constant is large enough. Unfortunately, the critical value of this coupling constant that we obtain is much too large for the mean field theory to be reliable and, hence, we cannot exclude the possibility that other states may arise at weaker coupling. Nevertheless, it is an instructive excercise that shows that ring-exchange processes, if large enough, may trigger a chiral spin liquid on their own.

In this section we explore of the possibility of breaking this symmetry spontaneously. Numerical works have studied examples where such scenarios arise in the Heisenberg model on the kagome lattice in the presence of second and third next nearest neighbor Heisenberg terms or Ising terms,\cite{He2014, Gong:2014} where they find suggestive evidence of a chiral spin liquid in certain regimes. Unfortunately, the flux attachment transformation summarized  in Sec. \ref{sec:flux-attachment-kagome} cannot be applied to next nearest neighbor Heisenberg terms. However, we have examined the case in the presence of just the next nearest neighbor Ising terms, using flux attachment methods and  we do not find the chiral spin liquid observed in the numerical work.\cite{He2014}

As a result we consider the effect of adding a chiral term on a bowtie in the kagome lattice which is written explicitly as follows
\beq
H_{ \rotatebox{90}{$\bowtie$} } = g \sum_{ \{ \rotatebox{90}{$\bowtie$} \} } 
\left[ {\bm S}_i \cdot ({\bm S}_j \times {\bm S}_k ) \right] \left[ {\bm S}_i \cdot ({\bm S}_l \times {\bm S}_m ) \right]
\label{eq:Bowtie_Hamiltonian}
\eeq
where the sum runs over all the bowties of the kagome lattice, $i$, $j$, and $k$ refer to the indices of the up triangle and $i$, $l$ and $m$ refer to the indices of the down triangle, with $i$ being the common site in the bowtie. 

The total Hamiltonian used in this section, can then be written as
\begin{align}
H_{tot} =&  H_{XXZ} + H_{ \rotatebox{90}{$\bowtie$}  }\nonumber\\
            =& H_{XXZ}+\sum_{\langle \triangle, \bigtriangledown \rangle } g \chi_{\triangle} \chi_{\bigtriangledown} 
\end{align}
where the $H_{XXZ}$ is the Hamiltonian  of the Heisenberg antiferromagnet on the kagome lattice, with anisotropy coupling $\lambda$,  defined in Eq. \eqref{eq:Heisenberg_Hamiltonian}, and where $\chi_{\triangle}$ and $\chi_{\bigtriangledown}$ are the chiralities over the up and down triangles (i.e. the sites of the two sublattices of the honeycomb lattice) and the sum runs over nearest-neighbor triangles of the kagome lattice (which correspond to the bowties). In what follows we will assume that we are either at the isotropic point or in the regime of $XY$ anisotropy (easy plane), i.e. $\lambda\leq 1$.

We now note that the bowtie terms of the Hamiltonian in Eq.\eqref{eq:Bowtie_Hamiltonian}, when expanded, can be expressed in terms of a ring-exchange term on the bowtie as follows
{\small
\beq
H_{\rotatebox{90}{$\bowtie$} } =  \frac{g}{2} \sum_{\{ \rotatebox{90}{$\bowtie$} \} } \bigg\{ & {\bm S}_i \cdot {\bm S}_i \left[ ({\bm S}_j \cdot {\bm S}_l ) ({\bm S}_k \cdot {\bm S}_m )
 - ({\bm S}_j \cdot {\bm S}_m ) ({\bm S}_k \cdot {\bm S}_l ) \right] \\
 + & {\bm S}_i \cdot {\bm S}_l \left[ ({\bm S}_j \cdot {\bm S}_m ) ({\bm S}_i \cdot {\bm S}_k ) 
 -  ({\bm S}_i \cdot {\bm S}_j ) ({\bm S}_k \cdot {\bm S}_m ) \right] \\	
 + & {\bm S}_i \cdot {\bm S}_m \left[ ({\bm S}_i \cdot {\bm S}_j ) ({\bm S}_k \cdot {\bm S}_l ) 	
 - ({\bm S}_j \cdot {\bm S}_l ) ({\bm S}_i \cdot {\bm S}_k ) \right] \bigg\}
 \label{eq:Ring_exchange}
\eeq
}
Ring exchange terms have been known to give rise to exotic dimer states in Heisenberg antiferromagnets.\cite{Sandvik-2007} Here, we will explore the possibility of such a term giving rise to a chiral spin liquid state.

Since the triangles of the kagome can be labelled by the sites of a honeycomb lattice on the centers of the triangles, we can regard the Hamiltonian of Eq.\eqref{eq:Bowtie_Hamiltonian} as a coupling between the chiralities on a honeycomb lattice. Although Eq. \eqref{eq:Ring_exchange} has a very complicated form, it can be simplified by using a Hubbard-Stratonovich (HS) transformation in terms of a scalar field $h({\bm r}, t)$ on the sites $\{ {\bm r} \}$ to the honeycomb sublattice of the triangles of the kagome lattice. Upon this transformation, the action of the full system, $XXZ$ and chirality couplings, becomes
\begin{align}
S&=S_{XXZ}\nonumber\\
&+\int dt \frac{1}{2g} \sum_{{\bm r, {\bm r}'}} h({\bm r}, t) K^{-1}({\bm r}, {\bm r}') h({\bm r}', t)\nonumber\\
& -\int dt \sum_{{\bm r}} h({\bm r}, t) \chi({\bm r},t)
\label{eq:HS-bowties}
\end{align}
where $K({\bm r}, {\bm r}')$ is the coordination (or connectivity)  matrix of the honeycomb lattice and $K^{-1}({\bm r}, {\bm r}')$ is its inverse. The HS field $h({\bm r},t)$ plays the role of the chirality field introduced in
Sec.\ref{sec:MF_chiral}, except that here it is a function of time and space.

We can now apply the flux-attachment transformation to a system whose action is given by Eq.\eqref{eq:HS-bowties}, and, as we did in the preceding sections, map this problem to a system of fermions on the kagome lattice coupled to a lattice Chern-Simons gauge field. However now they are also coupled to the HS fields $h({\bm r},t)$ in the same fashion as we coupled the fermions to the chiral operator in Sec. \ref{sec:MF_chiral}. 

We can now integrate out the fermions,  we obtain the following effective action
\begin{align}
S&=S_{XXZ}\nonumber\\
&+\int dt \frac{1}{2g} \sum_{{\bm r, {\bm r}'}} h({\bm r}, t) K^{-1}({\bm r}, {\bm r}') h({\bm r}', t)\nonumber\\
& +S_{\rm eff}[h({\bm r},t),A_\mu({\bm r},t)]
\label{eq:Stot}
\end{align}
where $S_{\rm eff}[h({\bm r},t),A_\mu({\bm r},t)]$ is the effective action of the fermions in a background chirality field $h({\bm r},t)$ (and which includes the lattice Chern-Simons term, as before).

We can now carry out a mean-field approximation by extremizing the action of Eq.\eqref{eq:Stot} with respect to the chirality field $h({\bm r},t)$, and to the gauge field $A_\mu$. Since we are working at zero external magnetic field, the mean field state for the gauge field is just the $(\pi, \pi, \pi)$ flux state and, hence, in the absence of any other interactions, we will naively have two species of massless Dirac fermions (as discussed in Sec.\ref{sec:Continuum}).
We will 
take the extremal HS field to have a time-independent value on each sublattice, ${\bar h}_\triangle$ and ${\bar h}_\bigtriangledown$, which obey the 
equations
\begin{equation}
{\bar h}_\triangle=3g \langle \chi_{\bigtriangledown} \rangle, \qquad
{\bar h}_\bigtriangledown=3g \langle \chi_{\triangle} \rangle
\end{equation}
where $\langle \chi({\bm r},t)\rangle$ is the expectation value of the chirality operator on each sublattice. If we further seek solutions that do not break the sublattice symmetry, we obtain the simple mean field equation for the chirality
\begin{equation}
{\bar h}=3g\;  \langle \chi \rangle
\end{equation}
and the critical value of the chirality coupling $g_c$ is given by the usual mean-field-theory relation
\begin{equation}
1=3g_c \frac{d \langle \chi \rangle}{dh}\Big|_{h=0}
\end{equation}
where $\frac{d \langle \chi \rangle}{dh}\Big|_{h=0}$ is the chirality susceptibility of the $XXZ$ model.

\begin{figure}
  \begin{center}
    \includegraphics[width=0.5\textwidth]{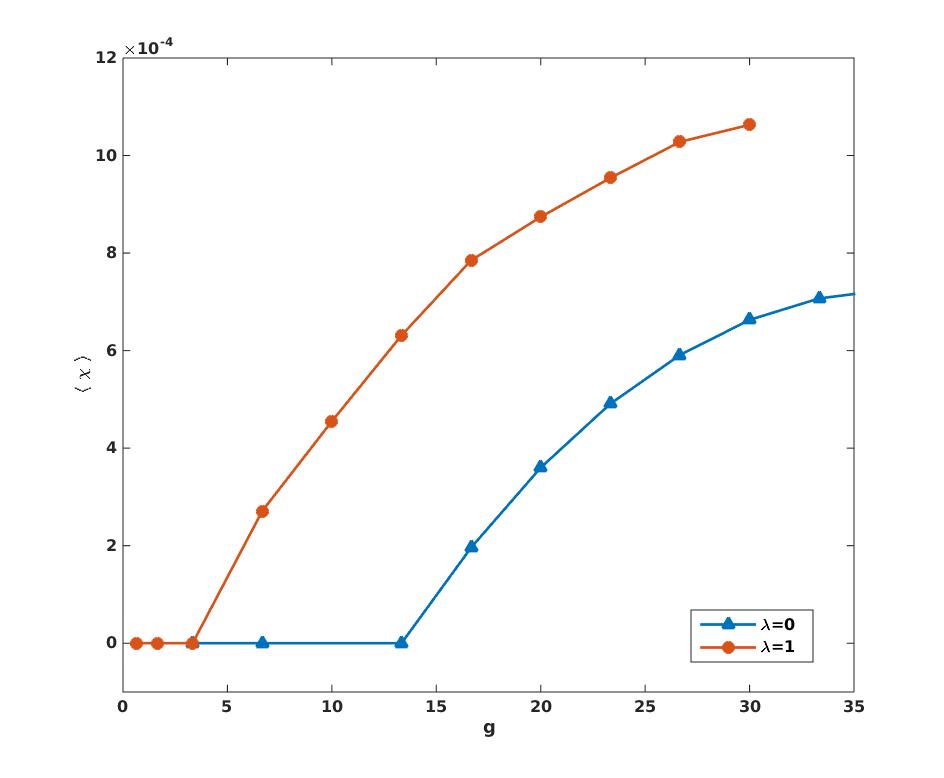}
  \end{center}
  \caption{(Color online) Expectation value of the chirality operator $\langle \chi \rangle$ plotted as a function of $g$ for $\lambda=0$, the $XY$ limit (full circles) and $\lambda/J=1$, the isotropic Heisenberg point (triangles). The mean-field theory critical values are $g_c \approx 13.3J$  in the $XY$ limit, $\lambda=0$, and $g_c \approx 3.4J$ at the isotropic Heisenberg point,  $\lambda/J=1$.}
  \label{fig:Chvsg}
\end{figure}

For 
$\lambda/J = 0$, we find that for values of $g \gtrsim 13.3J$, there exist non-vanishing solutions of the chirality parameter i.e. $h > 0$ as can be seen in Fig. \ref{fig:Chvsg}. In these cases, we end up with a non-zero chiral term similar to that of Eq. \eqref{eq:Chiral_Hamiltonian} and the
resultant phase would again be gapped and correspond to the chiral spin liquid discussed in the previous section. The critical value of $g$ reduces as one approaches the isotropic point. For $\lambda/J = 1$, the critical value is much smaller, $g_c \approx 3.4J$. Below this critical value of $g_c$, the value of $h$ that satisfy the mean-field consistency equations are $h= 0$. In this situation, we are back to the situation with just the $XXZ$ Heisenberg model and the resultant phase at half-filling would be gapless. The expectation values displaced in Fig \ref{fig:Chvsg} are quite small. The main reason for this is that each chirality operator has a term proportional to $S_z$. When all the sites are exactly at half-filling this terms is equal to zero (See Eq. \eqref{eq:Sz_map}) and the chiral expectation vanishes. In order to open up a gap, the densities have to be slightly shifted away from zero giving rise to a small non-zero chiral expectation value.

This leaves us with the question of what is the ground state of the $XXZ$ Heisenberg antiferromagnet on the kagome lattice for small $\lambda<1$ and $g<g_c$. Naively, we  would seem to predict that it is equivalent to a theory of two massless Dirac fermions which, on many grounds, cannot be the correct answer. In fact, L{\'o}pez, Rojo and one of us\cite{Lopez1994} found the same result in the $XY$ regime of the quantum Heisenberg antiferromagnet {\em on the square lattice} (which is not frustrated). These authors showed that the naive expectation is actually wrong and the fermions became massive by a process that can be represented as the exchange of Chern-Simons gauge bosons. Due to the stronger infrared behavior of the Chern-Simons gauge fields (compared with , e.g.,  Maxwell), this exchange term leads to an induced mass term for the Dirac fermions which is infrared finite (but linearly divergent in the ultraviolet). Most significantly the sign of the induced mass term leads to an extra Chern-Simons term which exactly cancelled the term introduced by flux attachment, leaving a parity-invariant Maxwell-type term as the leading contribution to the effective action. Furthermore, in 2+1 dimensions, a Maxwell term is known to be dual to a Goldstone boson. L{\'o}pez {\it et al.} concluded that the ground state  of the antiferromagnet on the square lattice in the $XY$ regime has long range order and that the Goldstone mode is just the Goldstone mode of the broken U(1) symmetry of this anisotropic regime. It should be apparent that in our case we can repeat the same line of argument almost verbatim which would suggest that in the $XY$ regime the ground state of the antiferromagnet on the kagome lattice should also have long range order with a broken U(1) symmetry. However, this conclusion is at variance with the best available numerical evidence which suggests, instead, that the ground state is $\mathbb{Z}_2$  spin liquid (of the Toric Code variety). The resolution of this issue is an open question.

In summary, this mean field theory predicts that beyond some critical value of the ring-exchange coupling constant $g$, which in this mean-field-theory is typically large, the system is in a chiral spin liquid state with a spontaneously broken time reversal invariance. However, below this critical value the mean field theory seemingly predicts that the Heisenberg antiferromagnet on the kagome lattice is in a phase with two species of gapless Dirac fermions. However this is not (and cannot be)  the end of the story. Indeed, the fermions are strongly coupled to the Chern-Simons gauge field which can (and should) change the story. In fact, in Ref. [\onlinecite{Lopez1994}] a similar result was found even in the case of a square lattice. A more careful analysis revealed that, in that case which is an unfrustrated system, the fermions acquired a mass in such a way that the total effective Chern-Simons gauge action vanished, resulting in a more conventional phase with a Goldstone mode. At present it is unclear what is the fate of the Dirac fermions in the case of the kagome lattice. In fact, most numerical data on the kagome antiferromagnet suggests that it is  a $\mathbb{Z}_2$ spin liquid. Whether a $\mathbb{Z}_2$ spin liquid  can be reproduced using our methods is an open problem.

\section{Conclusions}
\label{sec:Conclusions}

In this paper, we investigated the occurrence of chiral spin liquid phases in the nearest neighbor $XXZ$ Heisenberg Hamiltonian (with and without an external magnetic field) on the kagome lattice in the presence of various perturbations: a) a chirality symmetry breaking term, b)  Dzyaloshinski-Moriya interaction (only in the $XY$ limit), and c) ring-exchange interactions. At the mean-field level, we found that in the first two cases these interactions open up a gap in the spectrum and lead to phases with non-trivial Chern numbers (analogous to an integer quantum Hall state) in the $XY$ limit, $\frac{\lambda}{J} \lesssim 1$. When the effects of fluctuations are included, we find that these states actually correspond to fractional quantum Hall states for bosons with a spin Hall conductivity of $\sigma_{xy}^s = \frac{1}{2}$. This chiral spin liquid state survives for larger values of the chirality term but for larger values of the Dzyaloshinski-Moriya term, the chiral spin liquid state vanishes. Our results qualitatively agree with those obtained in a recent numerical study using the same model.\cite{Bauer-2014}

We also considered the effects of adding ring-exchange term on the bowties of the kagome lattice and found that, provided the coupling constant is larger than a critical value (which depends on parameters, e.g. the value of the Ising interaction), time-reversal symmetry is spontaneously broken and results in a topological state  similar chiral spin liquid state. However since the critical couplings that we find are rather large, ranging from $\frac{g}{J}\simeq 13.3$ in the $XY$ limit to $\frac{g}{J}\simeq 3.4$ at the isotropic point, we cannot exclude that other phases may also play a role. In particular, we have not explored the possible existence of topological phases with nematic order.\cite{Clark2013} 

In an earlier paper we showed that in the presence of a magnetic field, the nearest neighbor Heisenberg Hamiltonian gives rise to magnetization plateaus at $m=\frac{1}{3}$, $\frac{2}{3}$ and $\frac{5}{9}$ in the $XY$ limit.\cite{Kumar2014} Here, we found that some of these plateaus survive with the inclusion of the chirality and Dzyaloshinski-Moriya terms. In addition we also find another plateau at magnetization $m = \frac{1}{9}$ with a spin Hall conductivity $\sigma_{xy}^s = \frac{2}{3}$. 

In the absence of an external magnetic field, the flux attachment transformation that we use here, at the level of mean field theory, naively maps the kagome antiferromagnet onto a system of two species of massless Dirac fermions. Since this state   is not gapped,  the spectrum (and even the quantum numbers of the states) is not protected by the effects of fluctuations. Of all the fluctuations that are present, only the long range fluctuations of the Chern-Simons gauge field are (perturbatively) relevant. Indeed, this problem arises even in the simpler problem of the Heisenberg antiferromagnet on the square lattice, and  L{\'o}pez {\it et al.}\cite{Lopez1994} showed, using a non-trivial mapping,  that already at the one-loop level the spectrum changes from ``free'' massless Dirac fermions to the conventional N{\'e}el antiferromagnet with $XY$ anisotropy (easy plane). In Sec. \ref{sec:Continuum} we derived an effective field theory for the kagome antiferromagnet at zero field and, not surprisingly, found a state which naively has two species of massless Dirac fermions. A simple minded application of the same line of argument would also predict an easy-plane antiferromagnet which has a Goldstone mode (in the $XY$ regime). This however is not consistent with the best numerical data which shows no long range order but a topological $\mathbb{Z}_2$ state. How to reconcile these two scenarios is an open question which we are investigating. 

On the other hand, we should note that, contrary to the case of non-relativistic fermions, a theory of massless Dirac fermions coupled to a Chern-Simons gauge theory is non-trivial. While in  a massive phase this coupling should also amount to change in statistics, the massless case is much less understood. 
  In fact, the only case which a related problem is understood\cite{Giombi2012,Aharony2012,Maldacena2013} is the case in which the gauge fields have a gauge group $U(N)$ and the Chern-Simons action has level $k$.  In the limit in which $N\to \infty$ and $k\to \infty$ (with $\frac{N}{k}$ fixed), this problem maps onto a Wilson-Fisher fixed point of a scalar coupled to a Chern-Simons gauge theory with gauge group $U(k)$ at level $N$ (with the same ratio $\frac{N}{k}$). Away from this regime not much is known. In this large $N$ and large $k$ limit, the system remains conformally invariant (and hence critical). Our present understanding of the kagome antiferromagnets suggests that for small enough $N$ the system should become gapped and conformal symmetry should be spoiled.
   If the latter scenario is correct, then there should be a direct transition from (quite likely) a time-reversal invariant $\mathbb{Z}_2$ topological phase to a chiral spin liquid phase. If this were to hold  the quantum phase transition would most likely  be first order, although an exotic Landau-forbidden transition transition is also a possibility, perhaps of the deconfined quantum criticality type.\cite{Vishwanath-2003} In the latter case, the above cited recent  theories of conformal quantum field theories with Chern-Simons terms may be natural candidates for the field theory of such a quantum critical point.\cite{Giombi2012,Aharony2012,Maldacena2013}
  
\begin{acknowledgments}
We thank Bela Bauer, Andreas Ludwig and Ronny Thomale for stimulating discussions. 
EF thanks the KITP (and the Simons Foundation) and its IRONIC14 and ENTANGLED15 programs for support and hospitality, and to the Departamento de F{\'\i}sica (FCEyN), Universidad de Buenos Aires, for its hospitality.
This work was supported in part by the National Science Foundation by the grants PHY-1402971 at the University of Michigan (KS), DMR-1408713 at the University of Illinois (EF), PHY11-25915 at KITP (EF), and and by the U.S. Department of Energy, Division of Materials Sciences under Awards No. 
DE-FG02-07ER46453  through the Frederick
Seitz Materials Research Laboratory of the University of Illinois at Urbana-Champaign, and Ministry of Science and Technology (MINCyT, Argentina).
\end{acknowledgments}

%


\end{document}